\documentclass[amsmath, pra, twocolumn, showpacs, floatfix]{revtex4-1}
\usepackage{graphicx}
\usepackage{dsfont}

\begin{document}   

\title{Enhancement of the quadrupole interaction of an atom with guided light of an ultrathin optical fiber}
 
\author{Fam Le Kien,$^1$ Tridib Ray,$^2$ Thomas Nieddu,$^2$ Thomas Busch,$^1$ and S\'{i}le Nic Chormaic$^{2,3}$}

\affiliation{$^1$Quantum Systems Unit, Okinawa Institute of Science and Technology Graduate University, Onna, Okinawa 904-0495, Japan\\
$^2$Light-Matter Interactions Unit, Okinawa Institute of Science and Technology Graduate University, Onna, Okinawa 904-0495, Japan\\
$^3$School of Chemistry and Physics, University of KwaZulu-Natal, Durban, KwaZulu-Natal, 4001, South Africa}

\date{\today}

\begin{abstract}

We investigate the electric quadrupole interaction of an alkali-metal atom with guided light in the fundamental and higher-order modes of a vacuum-clad ultrathin optical fiber. We calculate the quadrupole Rabi frequency, the quadrupole oscillator strength, and their enhancement factors. In the example of a rubidium-87 atom, we study the dependencies of the quadrupole Rabi frequency on the quantum numbers of the transition, the mode type, the phase circulation direction, the propagation direction, the orientation of the quantization axis, the position of the atom, and the fiber radius. We find that the root-mean-square (rms) quadrupole Rabi frequency reduces quickly but the quadrupole oscillator strength varies slowly with increasing radial distance. We show that the enhancement factors of the rms Rabi frequency and the oscillator strength do not depend on any characteristics of the internal atomic states except for the atomic transition frequency. The enhancement factor of the oscillator strength can be significant even when the atom is far away from the fiber. We show that, in the case where the atom is positioned on the fiber surface, the oscillator strength for the quasicircularly polarized fundamental mode HE$_{11}$ has a local minimum at the fiber radius $a\simeq 107$ nm, and is larger than that for quasicircularly polarized higher-order hybrid modes, TE modes, and TM modes in the region $a<498.2$ nm. 

\end{abstract}

\maketitle

\section{Introduction}
\label{sec:intro}

Dipole-allowed optical transitions in atoms, ions, and molecules plays a key role in modern atomic, molecular, and optical physics \cite{DemtroderBook1}. The corresponding Rabi frequency is proportional to the intensity of the light field. Energy levels that are not connected to lower energy levels by dipole-allowed transitions are metastable states and have many applications ranging from precision clocks \cite{ClockRev15} to quantum gates \cite{RevIon}. Electric quadrupole transitions, on the other hand, are proportional to the gradient of the electric field and are less explored. Techniques to investigate non-dipole transitions have been explored theoretically and experimentally for atoms in free space \cite{Nilsen1977,Nilsen1978,Freedhoff1989,James1998,Kaler16,BDeb14,Ducloy16,Willitsch14,Gould09}, in evanescent fields \cite{Tojo2004,Tojo2005a,Tojo2005b}, near a dielectric microsphere \cite{Klimov1996}, near an ideally conducting cylinder \cite{Klimov2000}, and near plasmonic nanostructures \cite{Plasmon12,Shibata2017}. However, the difficulty in achieving large electric field gradients over a long distance makes the study of quadrupole transitions in an extended medium a challenging task.  

Ultrathin optical fibers \cite{TongNat03,review2016,review2017} allow tightly radially confined light to propagate over a long distance. Apart from a high intensity, the evanescent field that extends radially beyond the physical boundary of an ultrathin fiber also offers a large intensity gradient in the radial direction \cite{Tong04,fibermode}. The corresponding intensity gradient can be used to confine atoms near the surface of an ultrathin fiber \cite{fiber trap,Vetsch10,Goban12}. Furthermore, the higher-order modes of an ultrathin fiber \cite{Chormaic2015a,Fam2017a,Fam2017b} may also offer an azimuthal phase gradient. 

The aim of the present paper is to investigate the electric quadrupole interaction of an alkali-metal atom with guided light in the fundamental and higher-order modes of a vacuum-clad ultrathin optical fiber. We calculate the quadrupole Rabi frequency, the quadrupole oscillator strength, and their enhancement factors.
In the example of a rubidium-87 atom, we study the dependencies of these characteristics on the quantum numbers of the transition, the mode type, the phase circulation direction, the propagation direction, the orientation of the quantization axis, the position of the atom, and the fiber radius.
 
The paper is organized as follows. In Sec.~\ref{sec:quadrupole} we study the electric quadrupole interaction of an alkali-metal atom with an arbitrary monochromatic light field. In Sec.~\ref{sec:fiber} we examine the interaction of the atom with guided light of an ultrathin optical fiber and derive an expression for the enhancement factor of the quadrupole oscillator strength in terms of the fiber mode functions. In Sec.~\ref{sec:numerical} we present numerical results. Our conclusions are given in Sec.~\ref{sec:summary}.

\section{Quadrupole interaction of an atom with an arbitrary light field}
\label{sec:quadrupole}

\begin{figure}[tbh]
\begin{center}
  \includegraphics{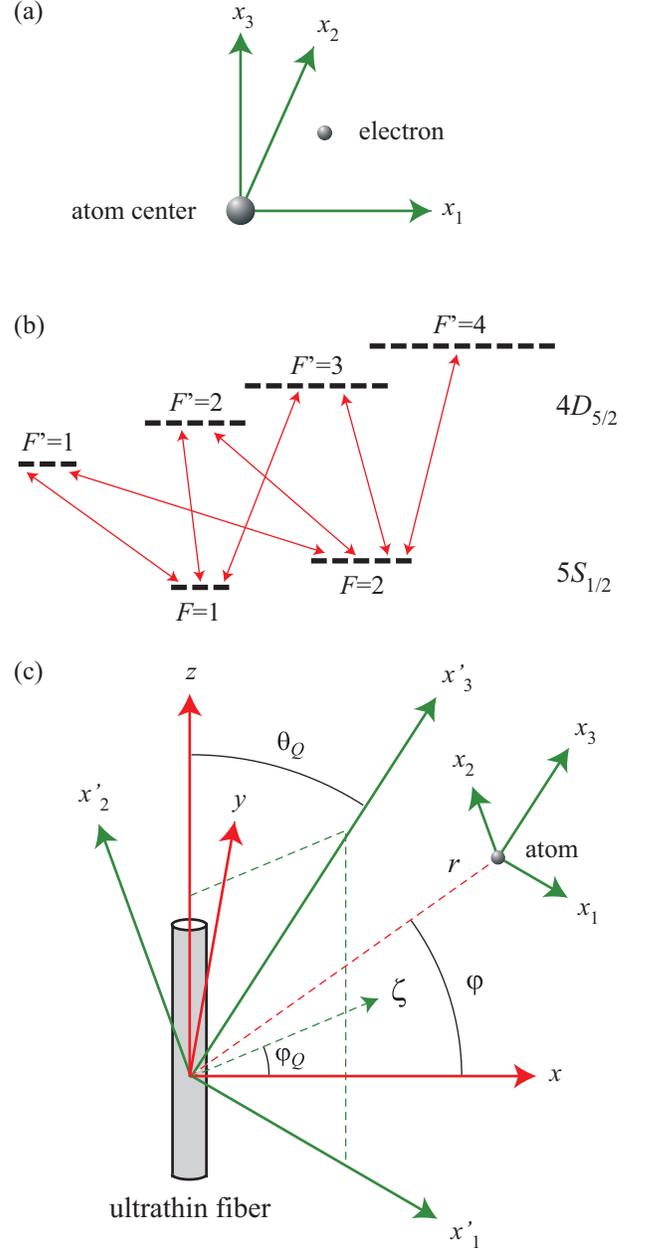}
 \end{center}
\caption{
(a) Local quantization coordinate system $\{x_1,x_2,x_3\}$ for an atom. 
(b) Schematic of the hfs levels of the $4D_{5/2}$  and $5S_{1/2}$ states of a rubidium-87 atom. 
(c) Atom in the vicinity of an ultrathin optical fiber with
the fiber-based Cartesian coordinate system $\{x,y,z\}$ and the corresponding cylindrical coordinate system $\{r,\varphi,z\}$.
}
\label{fig1}
\end{figure}

Consider an atom with a single valence electron interacting with an external optical field $\mathbf{E}$ through an electric quadrupole transition. We use Cartesian coordinates $\{x_1,x_2,x_3\}$ to describe the electric quadrupole and the internal states of the atom [see Fig.~\ref{fig1}(a)]. We assume that the center of mass of the atom is located at the origin $\mathbf{x}=0$ of this coordinate system. The electric quadrupole moment tensor $Q_{ij}$ of the atom, with $i,j=1,2,3$, is defined as 
\begin{equation}\label{a1}
Q_{ij}=e(3x_ix_j-R^2\delta_{ij}),
\end{equation}
where $x_i$ is the $i$th coordinate of the valence electron of the atom and $R=\sqrt{x_1^2+x_2^2+x_3^2}$ is the distance from the electron to the center of mass of the atom. The electric quadrupole interaction energy is \cite{Jackson}
\begin{equation}\label{a2}
W=-\frac{1}{6}\sum_{ij}Q_{ij}\frac{\partial E_j}{\partial x_i}\Big|_{\mathbf{x}=0},
\end{equation}
where the spatial derivatives of the field components  $E_j$ with respect to the coordinates $x_i$ are evaluated at the position $\mathbf{x}=0$ of the atom.
For simplicity, we neglect the effect of the surface-induced potential on the atomic energy
levels. This approximation is good when the atom is not too close to the fiber surface \cite{surface}.

We represent the field as $\mathbf{E}=(\boldsymbol{\mathcal{E}}e^{-i\omega t}+\boldsymbol{\mathcal{E}}^\ast e^{i\omega t})/2$,
where $\boldsymbol{\mathcal{E}}$ is the field amplitude and $\omega$ the field frequency.
Let $|e\rangle$ and $|g\rangle$ be upper and lower states of the atom, with energies $\hbar\omega_e$ and $\hbar\omega_g$, respectively.
In the interaction picture and the rotating wave approximation, the interaction Hamiltonian of the system can be written as
\begin{equation}\label{a3}
H_I=-\hbar\sum_{eg}\Omega_{ge} e^{-i(\omega-\omega_{eg}) t}\sigma_{eg}+\mathrm{H.c.},
\end{equation}
where $\omega_{eg}=\omega_e-\omega_g$ is the atomic transition frequency and
\begin{equation}\label{a4}
\Omega_{ge}=\frac{1}{12\hbar}\sum_{ij}\langle e|Q_{ij}|g\rangle\frac{\partial \mathcal{E}_j}{\partial x_i}
\end{equation}
is the Rabi frequency for the quadrupole transition between the states $|g\rangle$ and $|e\rangle$.

Consider the case of an alkali-metal atom with degenerate transitions between the magnetic sublevels $|g\rangle=|nFM\rangle$ and $|e\rangle=|n'F'M'\rangle$ [see Fig.~\ref{fig1}(b)]. 
Here, $n$ denotes the principal quantum number and also all additional quantum numbers not shown explicitly, 
$F$ is the quantum number for the total angular momentum  of the atom, and $M$ is the magnetic quantum number. 
The matrix elements $\langle n'F'M'|Q_{ij}|nFM\rangle$ of the quadrupole tensor operators $Q_{ij}$ are,
as shown in Appendix \ref{sec:tensor}, given as \cite{James1998}
\begin{eqnarray}\label{a5}
\lefteqn{\langle n'F'M'|Q_{ij}|nFM\rangle=3e u_{ij}^{(M'-M)}(-1)^{F'-M'}}\nonumber\\
&&\mbox{}\times
\begin{pmatrix}F' &2 &F \\-M' & M'-M& M\end{pmatrix}
\langle n'F'\|T^{(2)}\|nF\rangle,\qquad
\end{eqnarray}
where the matrices $u_{ij}^{(q)}$ with $q=-2,-1,0,1,2$ are given by Eqs.~\eqref{t12},
the array in the parentheses is a 3$j$ symbol, and the invariant factor $\langle n'F' \| T^{(2)}\|nF \rangle$ is the reduced matrix element
of the tensor operators $T_{q}^{(2)}=2(2\pi/15)^{1/2} R^2Y_{2q}(\theta,\varphi)$. Here, $Y_{lq}$ is a spherical harmonic function of degree $l$ and order $q$,
and $\theta$ and $\varphi$ are spherical angles in the spherical coordinates $\{R,\theta,\varphi\}$ associated with the Cartesian coordinates $\{x_1,x_2,x_3\}$.

It is clear from Eq.~\eqref{a5} that the selection rules for $F$ and $F'$ are $|F'-F|\le 2\le F'+F$, and the selection rule for $M$ and $M'$ is $|M'-M|\le 2$.
Since the tensor operators $T_{q}^{(2)}$ do not act on the nuclear spin degree of freedom, the dependence of the reduced matrix element 
$\langle n'F'\|T^{(2)}\|nF\rangle$ on $F$ and $F'$ may be factored out as \cite{tensor}
\begin{multline}\label{a6}
\langle n'F'\|T^{(2)}\|nF\rangle=(-1)^{J'+I+F}\\
\times\sqrt{(2F+1)(2F'+1)}\bigg\{\begin{array}{ccc}
F' &2 &F \\
J & I& J'
\end{array}\bigg\}
\langle n'J'\|T^{(2)}\|nJ\rangle,
\end{multline}
where $J$ is the quantum number for the total angular momentum of the electrons, $I$ is the nuclear spin quantum number,
and the array in the curly braces is a 6$j$ symbol. The selection rules for $J$ and $J'$ are $|J'-J|\le 2\le J'+J$.

Furthermore, since the tensor operators $T_{q}^{(2)}$ do not act on the electron spin degree of freedom, we have \cite{tensor}
\begin{multline}\label{a7}
\langle n'J'\|T^{(2)}\|nJ\rangle=(-1)^{L'+S+J}\\
\times\sqrt{(2J+1)(2J'+1)}\bigg\{\begin{array}{ccc}
J'&2 &J \\
L & S& L'
\end{array}\bigg\}
\langle n'L'\|T^{(2)}\|nL\rangle,
\end{multline}
where $L$ is the quantum number for the total orbital angular momentum of the electrons and $S$ the quantum number for the total spin of the electrons.
It follows from the addition of angular momenta that the quadrupole matrix elements may be nonzero only if  $|L'-L|\le 2\le L'+L$.
On the other hand, the parity of the tensor $T_{q}^{(2)}\propto Y_{2q}$ is even. Therefore, the quadrupole matrix elements may be nonzero only if $L$ and $L'$ have the same parity. Thus, the electric quadrupole transition selection rules for $L$ and $L'$ are $|L'-L|=0,2$ and $L'+L\ge 2$.
We note that, in the special case where $L=0$ and $L'=2$, we have $\langle n',L'=2\|T^{(2)}\|n,L=0\rangle=\sqrt{2/3}\langle n',L'=2|R^2|n,L=0\rangle$.

We now calculate the quadrupole Rabi frequency $\Omega_{ge}=\Omega_{FMF'M'}$, defined by Eq.~\eqref{a4}.
When we insert Eq.~\eqref{a5} into Eq.~\eqref{a4}, we obtain
\begin{equation}\label{a8}
\begin{split}
\Omega_{FMF'M'}&=\frac{e}{4\hbar}(-1)^{F'-M'}
\begin{pmatrix}F' &2 &F \\-M' & M'-M& M\end{pmatrix}\\
&\quad\times
\langle n'F'\|T^{(2)}\|nF\rangle\sum_{ij}u_{ij}^{(M'-M)}\frac{\partial \mathcal{E}_j}{\partial x_i}.
\end{split}
\end{equation}
In general, the Rabi frequency $\Omega_{FMF'M'}$ for the transition between the atomic states $|nFM\rangle$ and $|n'F'M'\rangle$ depends
on the relative orientation of the quantization axis $x_3$ with respect to the electric field vector $\boldsymbol{\mathcal{E}}$.

The root-mean-square (rms) Rabi frequency $\bar{\Omega}_{FF'}$ is given by the rule \cite{Shore} 
\begin{equation}\label{a9}
\bar{\Omega}_{FF'}^2=\sum_{MM'}|\Omega_{FMF'M'}|^2.
\end{equation}
We insert Eq.~\eqref{a8} into Eq.~\eqref{a9} and perform the summations over $M$ and $M'$. Then, we obtain
\begin{equation}\label{a10}
\bar{\Omega}_{FF'}^2=\frac{e^2}{80\hbar^2}|\langle n'F'\|T^{(2)}\|nF\rangle|^2
\sum_{q}\Big|\sum_{ij} u_{ij}^{(q)}\frac{\partial \mathcal{E}_j}{\partial x_i}\Big|^2.
\end{equation}

We note that Eqs.~\eqref{a8} and \eqref{a10} can be used for a monochromatic light field with an arbitrary space-dependent amplitude $\boldsymbol{\mathcal{E}}$. In the particular case of standing-wave laser fields, Eqs.~\eqref{a8} and \eqref{a10} reduce to the results of Ref.~\cite{James1998}.

We assume that the field is near to resonance with the atom, that is, $\omega\simeq\omega_0$,
where $\omega_0\equiv\omega_{eg}$. The oscillator strength $f_{FF'}$ can be calculated from the rms Rabi frequency $\bar{\Omega}_{FF'}$ 
by using the relation \cite{Shore}
\begin{equation}\label{a11}
\bar{\Omega}_{FF'}^{2}=\frac{e^2|\mathcal{E}|^2}{8\hbar m_e\omega_0}(2F+1)f_{FF'},
\end{equation}
where $m_{e}$ is the mass of an electron. 
This yields
\begin{equation}\label{a11a}
\begin{split}
f_{FF'}&=\frac{m_e\omega_0}{18\hbar e^2(2F+1)}\\
&\quad\times\sum_{MM'}\bigg|\sum_{ij}\langle n'F'M'|Q_{ij}|nFM\rangle
\frac{1}{\mathcal{E}}\frac{\partial\mathcal{E}_j}{\partial x_i}\bigg|^2.
\end{split}
\end{equation}
Equation \eqref{a11a} can be used for a monochromatic light field with an arbitrary space-dependent amplitude $\boldsymbol{\mathcal{E}}$. 
In the particular case where $\boldsymbol{\mathcal{E}}=\boldsymbol{\mathcal{E}}_0 e^{i\mathbf{K}\cdot\mathbf{x}}$ with $\boldsymbol{\mathcal{E}}_0$ and $\mathbf{K}$ being constant real or complex vectors, Eq.~\eqref{a11a} reduces to an expression that is in agreement with Refs.~\cite{Tojo2004,Tojo2005a,Tojo2005b}.

With the help of Eqs.~\eqref{a10} and \eqref{a11}, we find 
\begin{equation}\label{a12}
f_{FF'}=\frac{m_e\omega_0}{10\hbar}\frac{|\langle n'F'\|T^{(2)}\|nF\rangle|^2}{2F+1}
\sum_{q}\Big|\sum_{ij}u_{ij}^{(q)}\frac{1}{\mathcal{E}}\frac{\partial\mathcal{E}_j}{\partial x_i}\Big|^2.
\end{equation}
We emphasize that Eq.~\eqref{a12} can be used for an arbitrary monochromatic light field. 
Due to the summation over $M$ and $M'$ in Eq.~\eqref{a9}, the rms Rabi frequency $\bar{\Omega}_{FF'}$ and, consequently, the oscillator strength $f_{FF'}$ do not depend on the orientation of the quantization axis $x_3$. 
The quadrupole oscillator strength $f_{FF'}$, given by Eq.~\eqref{a12}, is a measure that characterizes the proportionality
of the rms Rabi frequency $\bar{\Omega}_{FF'}$ to the field magnitude $\mathcal{E}$ through Eq.~\eqref{a11}.
This measure depends on not only the quadrupole of the atom but also the normalized gradients of the field components. We note that, for atoms in free space, the oscillator strength can be interpreted as the ratio between the quantum-mechanical transition rate and the classical absorption rate of a single-electron oscillator with the same frequency \cite{Jackson,Shore}. However, this interpretation may not be valid for atoms in the vicinity of an object because 
the modifications of the transition rate are much more complicated than that of the Rabi frequency. 

According to expressions \eqref{a10} and \eqref{a12}, 
the dependencies of $\bar{\Omega}_{FF'}^2$ and $f_{FF'}$ on $F$ and $F'$ are included only in the factors 
$|\langle n'F'\|T^{(2)}\|nF\rangle|^2$ and $|\langle n'F'\|T^{(2)}\|nF\rangle|^2/(2F+1)$. These factors are determined
by the internal atomic states. They do not depend on the center-of-mass position of the atom and the parameters of the fiber.
They act as scaling factors for the dependencies on different $F$ and $F'$. Consequently, the shapes of the dependencies of $\bar{\Omega}_{FF'}^2$ and $f_{FF'}$
on the position of the atom and the radius of the fiber do not depend on the quantum numbers $F$ and $F'$.

We introduce the notations $\bar{\Omega}_{FF'}^{(0)}$ and $f_{FF'}^{(0)}$ for the rms Rabi frequency and oscillator strength
of an atom interacting with a plane-wave light field in free space via an electric quadrupole transition. 
According to \cite{James1998,Freedhoff1989,Tojo2005b}, we have 
\begin{equation}\label{a13}
\bar{\Omega}_{FF'}^{(0)2}=\frac{e^2k^2|\mathcal{E}|^2}{160\hbar^2}|\langle n'F'\|T^{(2)}\|nF\rangle|^2
\end{equation}
and
\begin{equation}\label{a14}
f_{FF'}^{(0)}=\frac{m_e\omega_0^3}{20\hbar c^2}\frac{|\langle n'F'\|T^{(2)}\|nF\rangle|^2}{2F+1}.
\end{equation}

The enhancements of the rms Rabi frequency and oscillator strength
in arbitrary light are characterized by the factors 
\begin{equation}\label{a15}
\begin{split}
\eta_{\mathrm{Rabi}}&=\frac{\bar{\Omega}_{FF'}}{\bar{\Omega}_{FF'}^{(0)}},\\
\eta_{\mathrm{osc}}&=\frac{f_{FF'}}{f_{FF'}^{(0)}}.
\end{split} 
\end{equation}
We find
\begin{equation}\label{a16}
\eta_{\mathrm{osc}}=\eta_{\mathrm{Rabi}}^2 = \frac{2}{k_0^2|\mathcal{E}|^2}\sum_{q}\Big|\sum_{ij} u_{ij}^{(q)}\frac{\partial \mathcal{E}_j}{\partial x_i}\Big|^2.
\end{equation}
It is clear from Eq.~\eqref{a16} that $\eta_{\mathrm{Rabi}}$ and $\eta_{\mathrm{osc}}$ are independent of the quantum numbers $F$ and $F'$.
Moreover, these factors do not depend on any characteristics of the atomic states except for the atomic transition frequency $\omega_0$. They are determined by the normalized spatial variations of the mode profile function $\boldsymbol{\mathcal{E}}$ at the frequency $\omega_0$.

We note that the oscillator strength $f_{JJ'}$ of the transition from a lower fine-structure level $|nJ\rangle$ to an upper fine-structure level $|n'J'\rangle$
of the atom may be obtained by summing up $f_{FF'}$ over all values of $F'$. 
The result is  
\begin{equation}\label{a17}
f_{JJ'}=\frac{m_e\omega_0}{10\hbar}\frac{|\langle n'J'\|T^{(2)}\|nJ\rangle|^2}{2J+1} 
\sum_{q}\Big|\sum_{ij}u_{ij}^{(q)}\frac{1}{\mathcal{E}}\frac{\partial\mathcal{E}_j}{\partial x_i}\Big|^2.
\end{equation}
In the case of an atom interacting with a plane-wave light field in free space, we have \cite{James1998,Freedhoff1989,Tojo2005b}  
\begin{equation}\label{a18}
f_{JJ'}^{(0)}=\frac{m_e\omega_0^3}{20\hbar c^2}\frac{|\langle n'J'\|T^{(2)}\|nJ\rangle|^2}{2J+1}.
\end{equation}
The relation between $f_{FF'}$ and $f_{JJ'}$ is \cite{Tojo2004,Tojo2005a,error}
\begin{equation}\label{a19}
f_{FF'}=(2F'+1)(2J+1)\begin{Bmatrix}F' &2 &F \\J & I& J'\end{Bmatrix}^2 f_{JJ'}.
\end{equation}
\section{Quadrupole interaction of an atom with guided light}
\label{sec:fiber}

We consider the electric quadrupole interaction between the atom and a guided light field of a vacuum-clad ultrathin optical fiber [see Fig.~\ref{fig1}(c)]. 
We assume that the fiber is a dielectric cylinder of radius $a$ and refractive index $n_1$ and is surrounded by an infinite background medium of refractive index $n_2$, where $n_2<n_1$.
We use Cartesian coordinates $\{x,y,z\}$, where $z$ is the coordinate along the fiber axis, and also cylindrical coordinates $\{r,\varphi,z\}$, where $r$ and $\varphi$ are the polar coordinates in the fiber transverse plane $xy$. 

We assume that the fiber supports the fundamental HE$_{11}$ mode and a few higher-order modes \cite{fiber books} 
in a finite bandwidth around the central frequency $\omega_0=\omega_e-\omega_g$ of the atom.
The theory of fiber guided modes is given in Ref.~\cite{fiber books} and is summarized in Appendix \ref{sec:guided}.
The propagation constant $\beta$ of a guided mode is determined by Eq.~\eqref{g1}.
We consider the class of quasicircularly polarized hybrid HE and EH modes, TE modes, and TM modes. 
A guided mode in this class can be labeled by an index $\mu=(\omega,N,f,p)$. 
Here, $\omega$ is the mode frequency, the notation $N=\mathrm{HE}_{lm}$, EH$_{lm}$, TE$_{0m}$, or TM$_{0m}$ stands for the mode type, with $l=1,2,\dots$ and $m=1,2,\dots$ being the azimuthal and radial mode orders, respectively, the index $f=+1$ or $-1$ denotes respectively the forward or backward propagation direction along the fiber axis $z$, and $p$ is the polarization index. The HE$_{lm}$ and EH$_{lm}$ modes are hybrid modes. For these modes, the azimuthal order is $l\not=0$,
and the index $p$ is equal to $+1$ or $-1$, indicating the counterclockwise or clockwise circulation direction of the helical phasefront. The TE$_{0m}$ and TM$_{0m}$ modes are transverse electric and magnetic modes. For these modes, the azimuthal mode order is $l=0$, the mode polarization is single, and the polarization index $p$ can be dropped. 

For a quasicircularly hybrid $\mathrm{HE}_{lm}$ or EH$_{lm}$ mode with the propagation direction $f$ and the phase circulation direction $p$, the field amplitude is \cite{fiber books,Fam2017a}
\begin{equation}\label{a20}
\boldsymbol{\mathcal{E}}=(e_r\hat{\mathbf{r}}+pe_\varphi\hat{\boldsymbol{\varphi}}+fe_z\hat{\mathbf{z}})e^{if\beta z+ipl\varphi},
\end{equation}
where $e_r$, $e_\varphi$, and $e_z$ are given by Eqs.~\eqref{g10} and \eqref{g11} for $\beta>0$ and $l>0$. 

For a TE$_{0m}$ mode with the propagation direction $f$, the field amplitude is \cite{fiber books,Fam2017a}
\begin{equation}\label{a21}
\boldsymbol{\mathcal{E}}=e_\varphi\hat{\boldsymbol{\varphi}}e^{if\beta z},
\end{equation}
where the only nonzero cylindrical component $e_\varphi$ is given by Eqs.~\eqref{g12} and \eqref{g13}.

For a TM mode with the propagation direction $f$, the field amplitude is \cite{fiber books,Fam2017a}
\begin{equation}\label{a22}
\boldsymbol{\mathcal{E}}=(e_r\hat{\mathbf{r}}+fe_z\hat{\mathbf{z}})e^{if\beta z},
\end{equation}
where the components $e_r$ and $e_z$ are given by Eqs.~\eqref{g14} and \eqref{g15} for $\beta>0$.
An important property of the mode functions of hybrid and TM modes is that the longitudinal
component $e_z$ is nonvanishing and in quadrature ($\pi/2$ out of phase) with the radial component $e_r$.

Quasilinearly polarized hybrid modes are linear superpositions of counterclockwise and clockwise quasicircularly polarized hybrid modes. The amplitude of the guided field in a quasilinearly polarized hybrid mode can be written in the form 
\begin{eqnarray}\label{a23}
\boldsymbol{\mathcal{E}}&=&\sqrt2[\hat{\mathbf{r}}e_r\cos (l\varphi-\varphi_{\mathrm{pol}})
+i\hat{\boldsymbol{\varphi}}e_\varphi\sin (l\varphi-\varphi_{\mathrm{pol}})\nonumber\\
&&\mbox{}+f\hat{\mathbf{z}}e_z\cos (l\varphi-\varphi_{\mathrm{pol}})]e^{if\beta z},\nonumber\\
\end{eqnarray}
where the phase angle $\varphi_{\mathrm{pol}}$ determines the orientation of the symmetry axes of the mode profile in the fiber transverse plane. In particular, the specific phase angle values $\varphi_{\mathrm{pol}}=0$ and $\pi/2$ define two orthogonal polarization profiles, one being symmetric with respect to the $x$ axis and the other being the result of the rotation of the first one by an angle of $\pi/2l$ in the fiber transverse plane $xy$.

In order to calculate the quadrupole Rabi frequency $\Omega_{FMF'M'}$ and the quadrupole oscillator strength $f_{FF'}$,
we need to transform the position and the field from the coordinate system $\{x,y,z\}$ to the coordinate system $\{x_1,x_2,x_3\}$. 
For this purpose, we translate the local coordinate system $\{x_1,x_2,x_3\}$ from the position of the atom to the origin
of the fiber-based coordinate system $\{x,y,z\}$. We denote the new coordinate system as $\{x'_1,x'_2,x'_3\}$.
Let $\theta_Q$ be the angle between the quantization axis $x'_3$ and the fiber axis $z$  [see Fig.~\ref{fig1}(c)].
Assume that the plane $(z,x'_3)$ intersects with the fiber transverse plane $xy$ at a line $\zeta$. Let $\varphi_Q$ be the azimuthal angle between $\zeta$ and $x$. Without loss of generality, we choose the axes $x_1$ and $x_2$ such that $x'_1$ is in the plane $(z,x'_3)$ and $x'_2$ is in the plane $(x,y)$. 
Then, the transformation from the Cartesian coordinates $\{x,y,z\}$ to the Cartesian coordinates $\{x'_1,x'_2,x'_3\}$ reads
\begin{eqnarray}\label{a24}
x'_1&=& (x\cos\varphi_Q+y\sin\varphi_Q)\cos\theta_Q-z\sin\theta_Q,\nonumber\\ 
x'_2&=& -x\sin\varphi_Q+y\cos\varphi_Q,\nonumber\\  
x'_3&=& (x\cos\varphi_Q+y\sin\varphi_Q)\sin\theta_Q+z\cos\theta_Q.
\end{eqnarray}
The inverse transformation reads
\begin{equation}\label{a25}
\begin{split}
x&=(x'_1\cos\theta_Q+x'_3\sin\theta_Q)\cos\varphi_Q-x'_2\sin\varphi_Q,\\ 
y&=(x'_1\cos\theta_Q+x'_3\sin\theta_Q)\sin\varphi_Q+x'_2\cos\varphi_Q,\\  
z&=-x'_1\sin\theta_Q+x'_3\cos\theta_Q.
\end{split}
\end{equation}
The relations between $\{x_1,x_2,x_3\}$ and $\{x'_1,x'_2,x'_3\}$ are $x_1'=x_1+x'_{1a}$, $x_2'=x_2+x'_{2a}$, and $x'_3=x_3+x'_{3a}$,
where $(x'_{1a},x'_{2a},x'_{3a})$ are the coordinates of the position of the atom in the system $\{x'_1,x'_2,x'_3\}$.
 
Meanwhile, the transformation from the components of the field vector $\boldsymbol{\mathcal{E}}$ in the Cartesian coordinate system $\{x,y,z\}$ to that in the Cartesian coordinate system $\{x_1,x_2,x_3\}$ is given by the equations
\begin{eqnarray}\label{a26}
\mathcal{E}_{x_1}&=& (\mathcal{E}_x\cos\varphi_Q+\mathcal{E}_y\sin\varphi_Q)\cos\theta_Q-\mathcal{E}_z\sin\theta_Q,\nonumber\\ 
\mathcal{E}_{x_2}&=& -\mathcal{E}_x\sin\varphi_Q+\mathcal{E}_y\cos\varphi_Q,\nonumber\\  
\mathcal{E}_{x_3}&=& (\mathcal{E}_x\cos\varphi_Q+\mathcal{E}_y\sin\varphi_Q)\sin\theta_Q+\mathcal{E}_z\cos\theta_Q.\qquad
\end{eqnarray}
The relations between the Cartesian-coordinate vector components $\mathcal{E}_x$ and $\mathcal{E}_y$ and the cylindrical-coordinate vector components $\mathcal{E}_r$ and $\mathcal{E}_\varphi$ are $\mathcal{E}_x=\mathcal{E}_r\cos\varphi-\mathcal{E}_\varphi\sin\varphi$ and $\mathcal{E}_y=\mathcal{E}_r\sin\varphi+\mathcal{E}_\varphi\cos\varphi$. With the help of the above transformations, we can easily calculate the quadrupole Rabi frequency $\Omega_{FMF'M'}$, the quadrupole oscillator strength $f_{FF'}$, and their enhancement factors $\eta_{\mathrm{osc}}$ and $\eta_{\mathrm{Rabi}}=\eta_{\mathrm{osc}}^{1/2}$ for guided light.

We now derive a simple analytical expression for the enhancement factor $\eta_{\mathrm{osc}}$ for quasicircularly hybrid HE and EH modes, TE modes, and TM modes. Since $\eta_{\mathrm{osc}}$ does not depend on the orientation of the quantization axis $x_3$, we use, without loss of generality, the fiber coordinate system $\{x,y,z\}$ as the quantization coordinate system, that is, we take $\{x_1,x_2,x_3\}$ such that $\{x'_1,x'_2,x'_3\}=\{x,y,z\}$. 
In addition, we assume that the atom is positioned on the positive side of the $x$ axis, that is, we set $\varphi=z=0$. 
Then, for a quasicircularly hybrid HE or EH mode, a TE mode, or a TM mode, we have
\begin{eqnarray}\label{a27}
&&\frac{\partial \mathcal{E}_1}{\partial x_1}=e'_r,
\quad \frac{\partial \mathcal{E}_2}{\partial x_1}=pe'_\varphi,
\quad \frac{\partial \mathcal{E}_3}{\partial x_1}=fe'_z,\nonumber\\
&&\frac{\partial \mathcal{E}_1}{\partial x_2}=\frac{p}{r}(il e_r-e_\varphi),
\, \frac{\partial \mathcal{E}_2}{\partial x_2}=\frac{1}{r}(il e_\varphi+e_r),
\, \frac{\partial \mathcal{E}_3}{\partial x_2}=\frac{fp}{r}il e_z,\nonumber\\
&&\frac{\partial \mathcal{E}_1}{\partial x_3}=if\beta e_r,
\quad \frac{\partial \mathcal{E}_2}{\partial x_3}=ifp\beta e_\varphi,
\quad \frac{\partial \mathcal{E}_3}{\partial x_3}=i\beta e_z,
\end{eqnarray}
where $e'_{r,\varphi,z}=\partial e_{r,\varphi,z}/\partial r$. 
When we insert Eqs.~\eqref{a27} into Eq.~\eqref{a16} and use Eqs.~\eqref{t12}, we find
\begin{eqnarray}\label{a28}
\eta_{\mathrm{osc}}&=&\frac{1}{k_0^2|\mathbf{e}|^2}\bigg[\Big|e'_r-\frac{1}{r}(ile_\varphi+e_r)\Big|^2+\Big|e'_\varphi+\frac{1}{r}(ile_r-e_\varphi)\Big|^2\nonumber\\
&&\mbox{}+|e'_z+i\beta_0 e_r|^2+\Big|\frac{l}{r}e_z+\beta_0 e_\varphi\Big|^2\nonumber\\
&&\mbox{}+\frac{1}{3}\Big|e'_r-2i\beta_0 e_z+\frac{1}{r}(ile_\varphi+e_r)\Big|^2\bigg],
\end{eqnarray}
where $\beta_0=\beta(\omega_0)$.
We can decompose $\eta_{\mathrm{osc}}$ as $\eta_{\mathrm{osc}}=\eta_{r}+\eta_{\varphi}+\eta_{z}+\eta_{\mathrm{mix}}$, where
\begin{eqnarray}\label{a29}
\eta_{r}&=&\frac{1}{k_0^2|\mathbf{e}|^2}\Big(\frac{4}{3}|e'_r|^2+|e'_\varphi|^2+|e'_z|^2\Big),\nonumber\\
\eta_{\varphi}&=&\frac{1}{k_0^2|\mathbf{e}|^2r^2}\Big(\frac{4}{3}|ile_\varphi+e_r|^2+|ile_r-e_\varphi|^2+l^2|e_z|^2\Big),\nonumber\\
\eta_{z}&=&\frac{\beta_0^2}{k_0^2}\Big(1+\frac{1}{3}\frac{|e_z|^2}{|\mathbf{e}|^2}\Big)
\end{eqnarray}
are the contributions from the field gradients in the $r$, $\phi$, and $z$ directions, respectively, and 
\begin{eqnarray}\label{a30}
\eta_{\mathrm{mix}}&=&
\frac{2}{3k_0^2|\mathbf{e}|^2r}\mathrm{Re}\Big[
2e'_r(ile_\varphi^*-e_r^*)+2i\beta_0e_z^*(re'_r+e_r)
\nonumber\\&&\mbox{}
-3e'_\varphi(ile_r^*+e_\varphi^*)-3i\beta_0r e'_z e_r^*+l\beta_0e_z e_\varphi^*\Big]
\end{eqnarray}
is a mixed term. In Eqs.~\eqref{a28}--\eqref{a30}, the mode functions and their spatial derivatives must be evaluated at the atomic transition frequency $\omega_0$. The expression for $\eta_z$ in Eqs.~\eqref{a29} indicates that $\eta_z$ is quadratically proportional to the propagation constant $\beta_0$ and increases with increasing magnitude $|e_z|/|\mathbf{e}|$ of the axial component of the polarization vector $\mathbf{e}/|\mathbf{e}|$.
Meanwhile, the expression for $\eta_\varphi$ in Eqs.~\eqref{a29} contains the factor $1/r^2$.
Due to this factor, $\eta_\varphi$ is small when $r$ and $a$ are large.
However, $\eta_\varphi$ may become significant when $r$ and $a$ are small.
It is clear that the expression for $\eta_\varphi$ in Eqs.~\eqref{a29} contains some terms with the coefficients proportional to $l$.
However, this expression also contains some terms with the coefficients independent of $l$. 
In addition, the mode function components $e_r$, $e_\varphi$, and $e_z$ depend on $l$ implicitly.
Consequently, the dependence of $\eta_\varphi$ on $l$ is complicated. In particular, $\eta_\varphi$ is not zero even for $l=0$, which corresponds to TE and TM modes.
We emphasize that, due to the summation over transitions with different magnetic quantum numbers and the cylindrical symmetry of the field in a quasicircularly hybrid HE or EH mode, a TE mode, or a TM mode, Eqs.~\eqref{a28}--\eqref{a30} remain valid for an arbitrary choice of the quantization axis $x_3$ and an arbitrary azimuthal position $\varphi$ of the atom.
For quasilinearly polarized hybrid modes, Eq.~\eqref{a27} and, consequently, Eqs.~\eqref{a28}--\eqref{a30} are not valid.

\section{Numerical results}
\label{sec:numerical}

In this section, we demonstrate the results of numerical calculations for the  
characteristics of an electric quadrupole transition of an atom interacting with a guided light field of an ultrathin optical fiber. As an example, we study the electric quadrupole transition between the ground state $5S_{1/2}$ 
and the excited state $4D_{5/2}$ of a rubidium-87 atom.
For this transition, we have $L'=2$, $J'=5/2$, $L=0$, $J=1/2$, $S=1/2$, and $I=3/2$.
The wavelength of the transition is $\lambda_0=516.5$ nm. The experimentally measured oscillator strength
of the transition $5S_{1/2}\to 4D_{5/2}$ in free space is 
$f_{JJ'}^{(0)}=8.06\times 10^{-7}$ \cite{Nilsen1978}. In our numerical calculations, we assume that the field is at exact resonance with the atom ($\omega=\omega_0$).

\begin{figure}[tbh]
\begin{center}
 \includegraphics{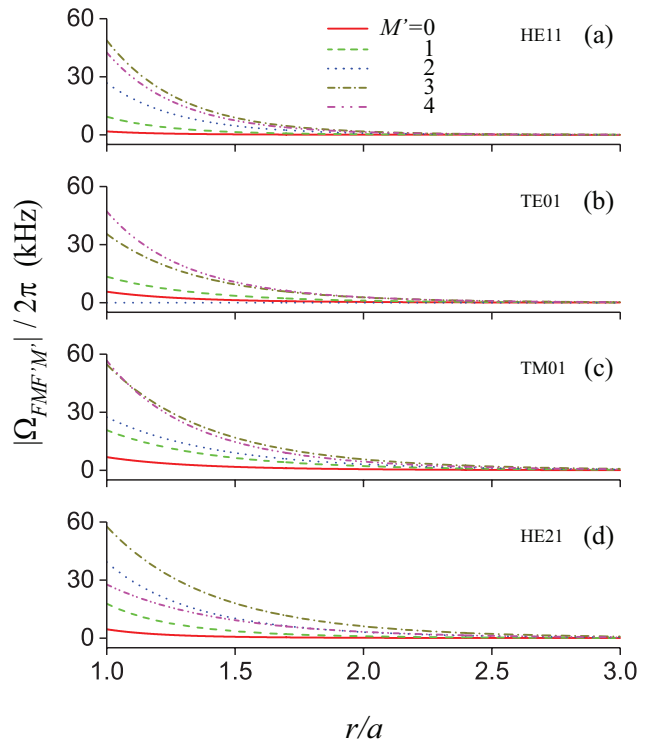}
 \end{center}
\caption{(Color online) Absolute value of the Rabi frequency $\Omega_{FMF'M'}$ for the quadrupole transition between the sublevel $M=2$ of the level $5S_{1/2}F=2$ and a sublevel $M'$ of the level $4D_{5/2}F'=4$ as a function of the radial distance $r$ for different magnetic quantum numbers $M'=0,1,2,3,4$ and different guided mode types $N=$ HE$_{11}$, TE$_{01}$, TM$_{01}$, and HE$_{21}$. The fiber radius is $a=280$ nm. The wavelength of the atomic transition is $\lambda_0=516.5$ nm. The refractive indices of the fiber and the vacuum cladding are $n_1=1.4615$ and $n_2=1$, respectively. The power of the guided light field is 10 nW. The field propagates in the $+z$ direction. The hybrid modes are counterclockwise quasicircularly polarized. The quantization axis is $x_3=z$. The azimuthal angle for the position of the atom in the fiber 
cross-sectional $xy$ plane is arbitrary.}
\label{fig2}
\end{figure}

We plot in Fig.~\ref{fig2} the absolute value of the Rabi frequency $\Omega_{FMF'M'}$ as a function of the radial distance $r$ for the transitions between a lower sublevel $|FM\rangle$ and different upper sublevels $|F'M'\rangle$
via the interaction with different guided modes $N=$ HE$_{11}$, TE$_{01}$, TM$_{01}$, and HE$_{21}$. 
For the calculations of this figure, we choose the quantization axis $x_3=z$.
We observe that $|\Omega_{FMF'M'}|$ reduces quickly with increasing $r$. The steep slope in the radial dependence of $|\Omega_{FMF'M'}|$ is a manifestation of the evanescent-wave behavior of the guided field outside the fiber. It is clear from Fig.~\ref{fig2} that $|\Omega_{FMF'M'}|$ depends
on the magnetic quantum numbers and the guided mode type. The dotted blue curve in Fig.~\ref{fig2}(b), which stands for the case of the upper sublevel $M'=2$ and the TE mode, is zero. This means that the TE mode does not interact with the quadrupole transition between the sublevels $|5S_{1/2},F=2, M=2\rangle$ and $|4D_{5/2},F'=4, M'=2\rangle$ for the quantization axis $x_3=z$. The vanishing of this interaction is a consequence of the properties of the TE mode, the quadrupole operator $Q_{ij}$, and the internal atomic states.

\begin{figure}[tbh]
\begin{center}
 \includegraphics{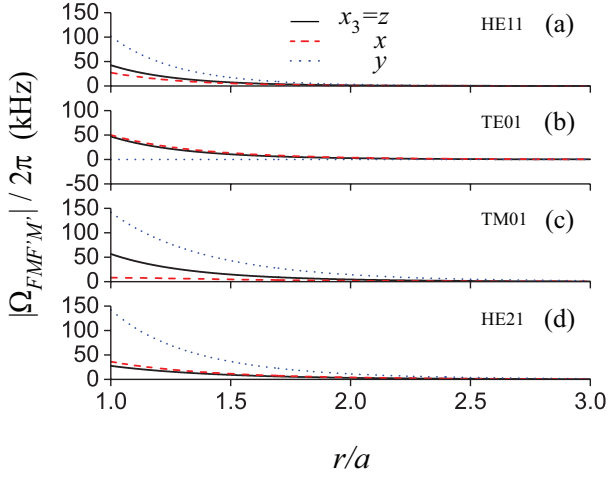}
 \end{center}
\caption{(Color online) Radial dependencies of the absolute value of the Rabi frequency $\Omega_{FMF'M'}$ for the quadrupole transition between the sublevels $|F=2,M=2\rangle$ and $|F'=4,M'=4\rangle$ for different choices of the quantization axis $x_3$ and different guided modes. The atom is positioned on the positive side of the $x$ axis ($\varphi=0$) and the hybrid modes are counterclockwise quasicircularly polarized. Other parameters are as for Fig.~\ref{fig2}.}
\label{fig3}
\end{figure}

The Rabi frequency $\Omega_{FMF'M'}$ for the transition between the sublevels $|FM\rangle$ and $|F'M'\rangle$ depends on the relative orientation of the quantization axis $x_3$ with respect to the fiber axis $z$. In order to illustrate this dependence, we plot in Fig.~\ref{fig3} the radial dependencies of the absolute value of the Rabi frequency $\Omega_{FMF'M'}$ 
for the quadrupole transition between the sublevels $|F=2,M=2\rangle$ and $|F'=4,M'=4\rangle$ for different choices of
the quantization axis, namely $x_3=z$, $x$, and $y$. 
We observe that $\Omega_{FMF'M'}$ strongly depends on the orientation of $x_3$. 
In the case of the HE$_{11}$, TM$_{01}$, and HE$_{21}$ modes, the absolute value
$|\Omega_{FMF'M'}|$ for $x_3=y$ [see the dotted blue curves in Figs.~\ref{fig3}(a), \ref{fig3}(c), and \ref{fig3}(d)] is larger than for  $x_3=z$ and $x_3=x$ [see the solid black and dashed red curves in Figs.~\ref{fig3}(a), \ref{fig3}(c), and \ref{fig3}(d)]. However, in the case of the TE$_{01}$ mode, we have $|\Omega_{FMF'M'}|=0$ for $x_3=y$ [see the dotted blue curve in Fig.~\ref{fig3}(b)].
The vanishing of this interaction is a consequence of the properties of the TE mode, the quadrupole operator $Q_{ij}$, and the internal atomic states.

\begin{figure}[tbh]
\begin{center}
 \includegraphics{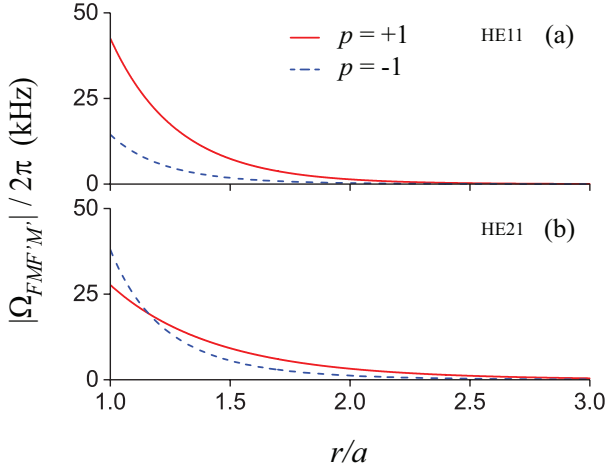}
 \end{center}
\caption{(Color online) Radial dependencies of the absolute value of the Rabi frequency $\Omega_{FMF'M'}$ for the opposite phase circulation directions $p=\pm1$ of the circularly polarized hybrid modes HE$_{11}$ and HE$_{21}$. 
The lower and upper levels of the transition are $|F=2,M=2\rangle$ and $|F'=4,M'=4\rangle$ and the quantization axis is $x_3=z$. Other parameters are as for Fig.~\ref{fig2}.}
\label{fig4}
\end{figure}

\begin{figure}[tbh]
\begin{center}
 \includegraphics{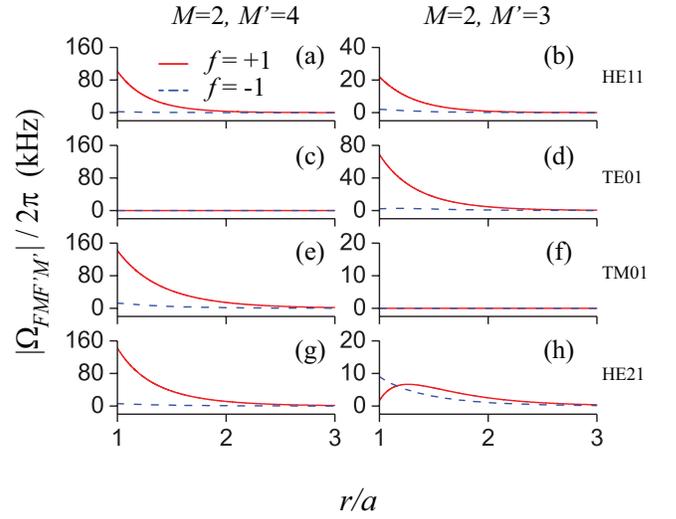}
 \end{center}
\caption{(Color online) Radial dependencies of the absolute value of the Rabi frequency $\Omega_{FMF'M'}$ for the opposite propagation directions $f=\pm1$ of different guided modes. The lower and upper levels of the transition are  
$|F=2,M=2\rangle$ and $|F'=4,M'=4\rangle$ (left column) and $|F=2,M=2\rangle$ and $|F'=4,M'=3\rangle$ (right column). 
The quantization axis is $x_3=y$, the atom is positioned on the positive side of the $x$ axis, and the hybrid modes are counterclockwise quasicircularly polarized. Other parameters are as for Fig.~\ref{fig2}.}
\label{fig5}
\end{figure}

We plot in Figs.~\ref{fig4} and \ref{fig5} the radial dependencies of $|\Omega_{FMF'M'}|$ 
for the opposite phase circulation directions $p=\pm1$ and the opposite propagation directions $f=\pm1$. 
These figures show that, depending on the orientation of the quantization axis, the mode type, and the transition type, 
$|\Omega_{FMF'M'}|$ may depend on $p$ and $f$. 
The dependence of $|\Omega_{FMF'M'}|$ on $f$ is related to the spin-orbit coupling of light \cite{Zeldovich,Bliokh review,Bliokh2014a,Bliokh2014b,Banzer review2015,Bliokh2015,Bliokh review2015}. It has been shown that, due to the spin-orbit coupling of light, spontaneous emission and scattering from an atom with a circular dipole near a nanofiber can be asymmetric with respect to the opposite propagation
directions along the fiber axis \cite{Fam2014,Petersen2014,Mitsch2014b,AtomArray,Sayrin2015b,Lodahl2017,Fam2017spon}.
We note that we have $|\Omega_{FMF'M'}|=0$ for both directions $f=\pm1$ in Figs.~\ref{fig5}(c) and \ref{fig5}(f). 
The vanishing of the quadrupole transitions in the cases of these figures is a consequence of the properties of the guided field, the quadrupole operator, and the internal atomic states.

\begin{figure}[tbh]
\begin{center}
 \includegraphics{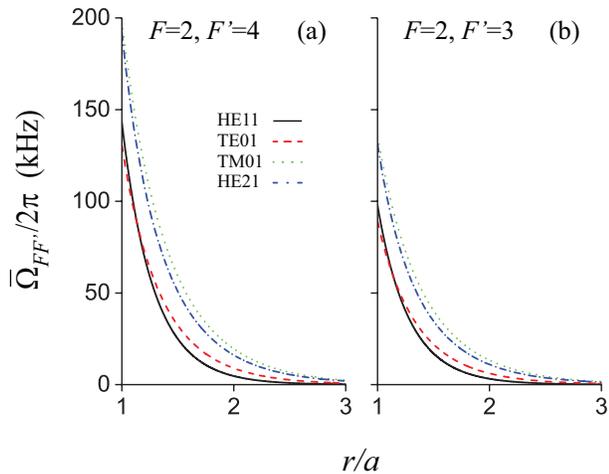}
 \end{center}
\caption{(Color online) Radial dependencies of the rms Rabi frequency $\bar{\Omega}_{FF'}$ for different guided modes.
The hfs levels are $F=2$ and $F'=4$ in (a) and $F=2$ and $F'=3$ in (b).
The hybrid modes are quasicircularly polarized and the quantization axis is arbitrary. Other parameters are as for Fig.~\ref{fig2}.}
\label{fig6}
\end{figure}

\begin{figure}[tbh]
\begin{center}
 \includegraphics{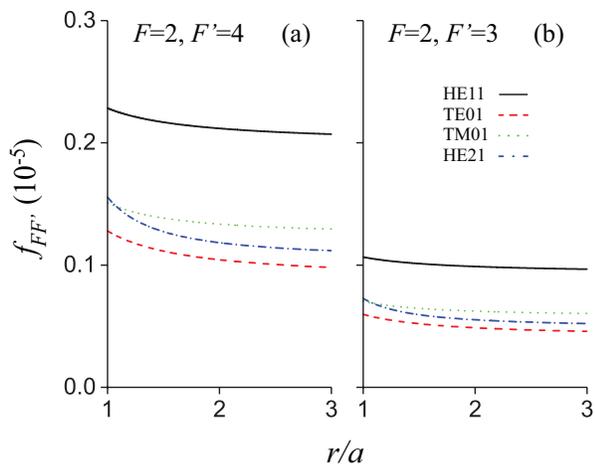}
 \end{center}
\caption{(Color online) Radial dependencies of the oscillator strength $f_{FF'}$ for different guided modes. Parameters used are as for Fig.~\ref{fig6}.}
\label{fig7}
\end{figure}

We plot in Figs.~\ref{fig6} and \ref{fig7} the radial dependencies of the rms Rabi frequency $\bar{\Omega}_{FF'}$ and the oscillator strength $f_{FF'}$ of the atom. 
As already pointed out in Sec.~\ref{sec:quadrupole}, due to the summation over transitions with different magnetic quantum numbers, $\bar{\Omega}_{FF'}$ and $f_{FF'}$ do not depend on the relative orientation of the quantization axis $x_3$ with respect to fiber axis $z$. Figures \ref{fig6} and \ref{fig7} show that $\bar{\Omega}_{FF'}$ and $f_{FF'}$ achieve their largest values at $r/a=1$. We observe that $\bar{\Omega}_{FF'}$ reduces quickly and $f_{FF'}$ decreases slowly  with increasing $r$. 
We note that the shapes of the curves in Figs.~\ref{fig6}(a) and \ref{fig7}(a), where $F=2$ and $F'=4$, are the same as the shapes of the corresponding curves in Figs.~\ref{fig6}(b) and \ref{fig7}(b), where $F=2$ and $F'=3$. The difference between these curves is given by a scaling factor [see Eqs.~\eqref{a10} and \eqref{a12}]. 

Figures \ref{fig6} and \ref{fig7} show that the rms Rabi frequency $\bar{\Omega}_{FF'}$ and the oscillator strength $f_{FF'}$ depend on the mode type. Comparison between the curves for different modes shows that, for the parameters of the figures,   
the oscillator strength $f_{FF'}$ for the fundamental mode HE$_{11}$ (see the solid black curve in Fig.~\ref{fig7}) is the largest, while the corresponding rms Rabi frequency $\bar{\Omega}_{FF'}$ (see the solid black curve in Fig.~\ref{fig6}) is the smallest or the second smallest. The contrast between these relations is due to the fact that the rms Rabi frequency $\bar{\Omega}_{FF'}$ is proportional to the product of the oscillator strength $f_{FF'}$ and the electric field intensity $|\mathcal{E}|^2$ [see Eq.~\eqref{a11}]. 
Outside the fiber, for a given power, the magnitude of the intensity of the field in the fundamental mode is smaller than that in other modes \cite{Fam2017a}.

\begin{figure}[tbh]
\begin{center}
 \includegraphics{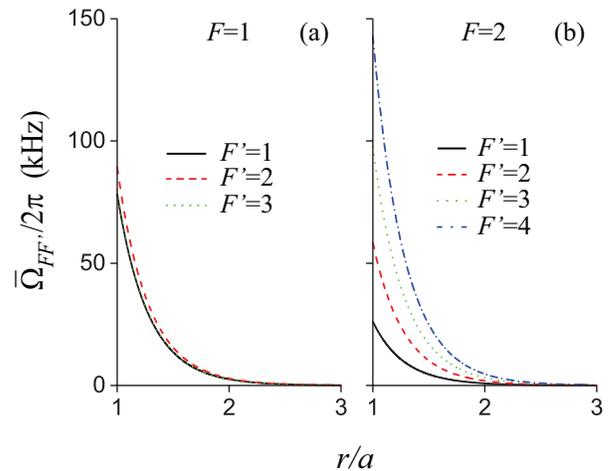}
 \end{center}
\caption{(Color online) Radial dependencies of the rms Rabi frequency $\bar{\Omega}_{FF'}$ of the atom interacting with the fundamental mode
HE$_{11}$ via the quadrupole transitions between different pairs of hfs levels $F$ and $F'$. The hybrid modes are quasicircularly polarized and the quantization axis is arbitrary. Other parameters are as for Fig.~\ref{fig2}.}
\label{fig8}
\end{figure}

\begin{figure}[tbh]
\begin{center}
 \includegraphics{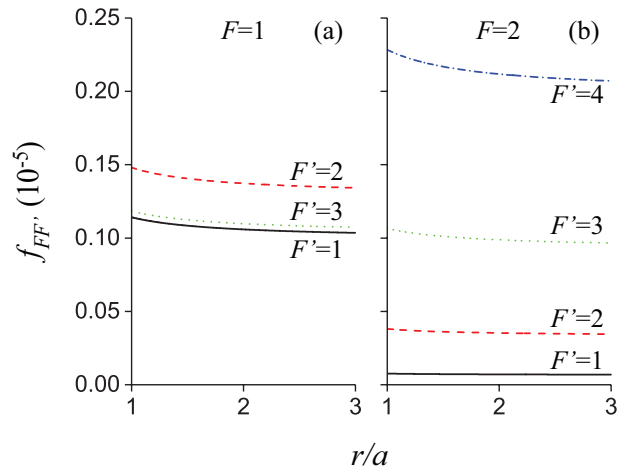}
 \end{center}
\caption{(Color online) Radial dependencies of the oscillator strength $f_{FF'}$ of the atom interacting with the fundamental mode
HE$_{11}$ via the quadrupole transitions between different pairs of hfs levels $F$ and $F'$. Parameters used are as for Fig.~\ref{fig8}.}
\label{fig9}
\end{figure}

We show in Figs.~\ref{fig8} and \ref{fig9} the radial dependencies of the rms Rabi frequency $\bar{\Omega}_{FF'}$ and the oscillator strength $f_{FF'}$ of the atom interacting with the fundamental mode HE$_{11}$ via the quadrupole transitions between different pairs of hfs levels $F$ and $F'$ of the ground state $5S_{1/2}$ and the excited stated $4D_{5/2}$. We observe from the figures that the curves for different pairs of $F$ and $F'$ have the same shape, that is, the curves for different pairs of $F$ and $F'$ are different from each other just by a scaling factor [see Eqs.~\eqref{a10} and \eqref{a12}]. Comparison between the curves show that the rms Rabi frequency and the oscillator strength are largest and smallest for the transitions between levels $F=2$ and $F'=4$ and between levels $F=2$ and $F'=1$, respectively. We note from Figs.~\ref{fig8}(a) and \ref{fig9}(a) that the transitions between levels $F=1$ and $F'=1$ and between levels $F=1$ and $F'=3$ have almost the same $\bar{\Omega}_{FF'}$ and the same $f_{FF'}$. 

\begin{figure}[tbh]
\begin{center}
 \includegraphics{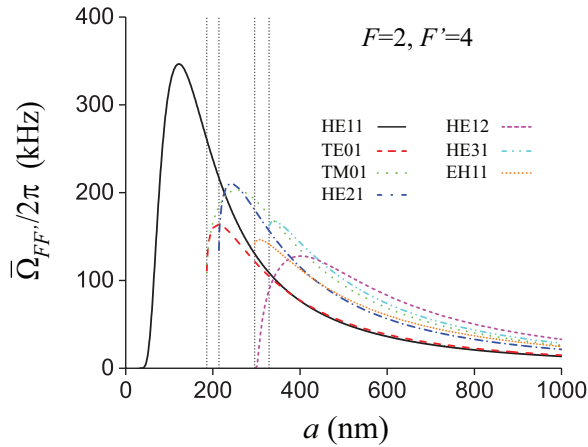}
 \end{center}
\caption{(Color online) Rms Rabi frequency $\bar{\Omega}_{FF'}$ as a function of the fiber radius $a$ for different guided modes. The atom is positioned on the fiber surface. 
The hybrid modes are quasicircularly polarized and the quantization axis is arbitrary. Other parameters are as for Fig.~\ref{fig2}. The vertical dotted lines indicate the positions of the cutoffs for higher-order modes.}
\label{fig10}
\end{figure}

\begin{figure}[tbh]
\begin{center}
 \includegraphics{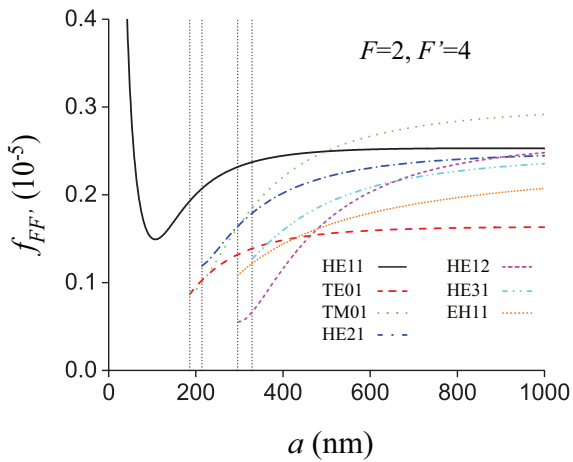}
 \end{center}
\caption{(Color online) Oscillator strength $f_{FF'}$ as a function of the fiber radius $a$ for different guided modes. Parameters used are as for Fig.~\ref{fig10}.}
\label{fig11}
\end{figure}

We show in Figs.~\ref{fig10} and \ref{fig11} the rms Rabi frequency $\bar{\Omega}_{FF'}$ and the oscillator strength $f_{FF'}$
as functions of the fiber radius $a$. We observe from Fig.~\ref{fig10} that the rms Rabi frequency $\bar{\Omega}_{FF'}$ first increases and then decreases with increasing $a$. It is clear from this figure that $\bar{\Omega}_{FF'}$ for different guided modes have different maxima at different values of $a$. We observe from Fig.~\ref{fig11} that, for the fundamental mode HE$_{11}$, the oscillator strength $f_{FF'}$ has a local minimum at $a\simeq 107$ nm. Meanwhile, for the higher-order modes, $f_{FF'}$ increases with increasing $a$.  In the region  $a<498.2$ nm, $f_{FF'}$ for the HE$_{11}$ mode is larger than that for  higher-order modes. When $a$ is in the region from $498.2$ nm to 1000 nm, $f_{FF'}$ for the TM$_{01}$ mode is lager than that for other modes. 

The increase of $f_{FF'}$ for the HE$_{11}$ and higher-order modes with increasing $a$ in the region of large $a$ is a consequence of the fact that expression \eqref{a12} for $f_{FF'}$ contains the terms that are proportional to the gradients $\partial{\mathcal{E}_{x,y,z}}/\partial z$ of the field amplitudes $\mathcal{E}_{x,y,z}$ in the direction of the fiber axis $z$. These gradients are proportional to the propagation constant $\beta$, which increases with increasing fiber radius $a$ \cite{fiber books,Fam2017a}. The decrease of $f_{FF'}$ with increasing $a$ in the region of small $a$ for the HE$_{11}$ mode (see the solid black curve in Fig.~\ref{fig11}) is a result of the changes in the structure of the field. The initial decrease and the subsequent increase lead to the occurrence of a minimum in the dependence of $f_{FF'}$ on $a$ in the case of the HE$_{11}$ mode (see the solid black curve in Fig.~\ref{fig11}).

\begin{figure}[tbh]
\begin{center}
 \includegraphics{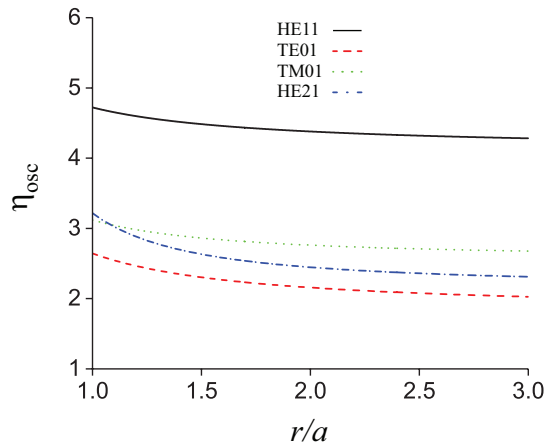}
 \end{center}
\caption{(Color online) Radial dependencies of the oscillator-strength enhancement factor $\eta_{\mathrm{osc}}$ for different guided modes. The hybrid modes are quasicircularly polarized and the quantization axis is arbitrary. 
Other parameters are as for Fig.~\ref{fig2}.}
\label{fig12}
\end{figure}

We plot in Fig.~\ref{fig12} the radial dependencies of the oscillator-strength enhancement factor $\eta_{\mathrm{osc}}$ for different guided modes. It is clear from the figure that $\eta_{\mathrm{osc}}$ achieves its largest values at $r/a=1$. We see that $\eta_{\mathrm{osc}}$ reduces slowly with increasing radial distance $r$. This result means that, despite the evanescent wave behavior, the enhancement factor $\eta_{\mathrm{osc}}$ can be significant even when the atom is far away from the fiber. The reason is that the oscillator strength $f_{FF'}$ and consequently the enhancement factor $\eta_{\mathrm{osc}}$ are determined by not the field amplitude but the ratio between the field gradient and the field amplitude. 
 
\begin{figure}[tbh]
\begin{center}
 \includegraphics{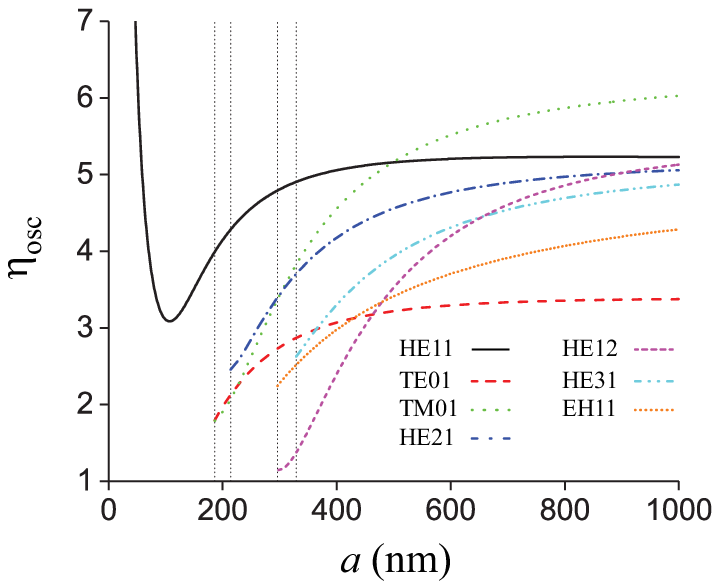}
 \end{center}
\caption{(Color online) Oscillator-strength enhancement factor $\eta_{\mathrm{osc}}$ as a function of the fiber radius $a$ for different guided modes. The atom is positioned on the fiber surface. The hybrid modes are quasicircularly polarized and the quantization axis is arbitrary. Other parameters are as for Fig.~\ref{fig2}. The vertical dotted lines indicate the positions of the cutoffs for higher-order modes.
}
\label{fig13}
\end{figure}

We show in Fig.~\ref{fig13} the oscillator-strength enhancement factor $\eta_{\mathrm{osc}}$ as a function of the fiber radius $a$ for different guided modes. 
Similar to the oscillator strength $f_{FF'}$, the enhancement factor $\eta_{\mathrm{osc}}$ for the fundamental mode HE$_{11}$ has a local minimum at the fiber radius $a\simeq 107$ nm, and is larger than that for higher-order modes in the region $a<498.2$ nm. Meanwhile, the enhancement factor $\eta_{\mathrm{osc}}$ for higher-order modes monotonically increases with increasing $a$. When $a$ is in the region from $498.2$ nm to 1000 nm, the factor $\eta_{\mathrm{osc}}$ for the TM$_{01}$ mode is lager than that for other modes.

\begin{figure}[tbh]
\begin{center}
 \includegraphics{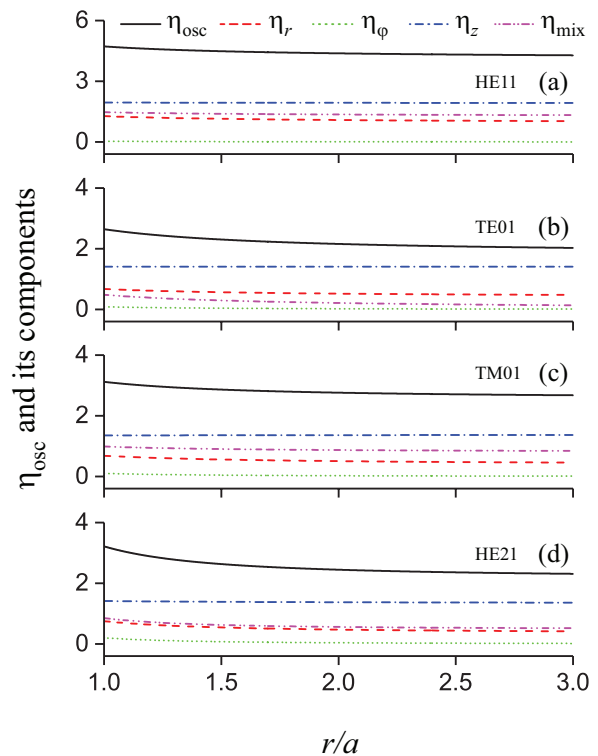}
 \end{center}
\caption{(Color online) Radial dependencies of the oscillator-strength enhancement factor $\eta_{\mathrm{osc}}$ and its components $\eta_r$, $\eta_\varphi$, $\eta_z$, and $\eta_{\mathrm{mix}}$ for different guided modes. The hybrid modes are quasicircularly polarized and the quantization axis is arbitrary. 
Other parameters are as for Fig.~\ref{fig2}.}
\label{fig14}
\end{figure}

\begin{figure}[tbh]
\begin{center}
 \includegraphics{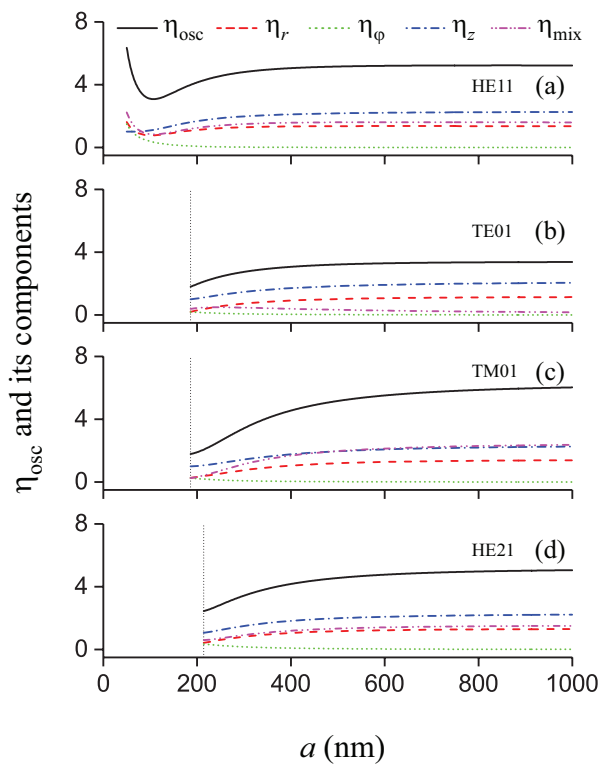}
 \end{center}
\caption{(Color online) Oscillator-strength enhancement factor $\eta_{\mathrm{osc}}$ and its components $\eta_r$, $\eta_\varphi$, $\eta_z$, and $\eta_{\mathrm{mix}}$ as functions of the fiber radius $a$ for different guided modes. The atom is positioned on the fiber surface. The hybrid modes are quasicircularly polarized and the quantization axis is arbitrary. Other parameters are as for Fig.~\ref{fig2}. The vertical dotted lines indicate the positions of the cutoffs for higher-order modes.
}
\label{fig15}
\end{figure}

According to Eq.~\eqref{a29}, the oscillator-strength enhancement factor $\eta_{\mathrm{osc}}$ can be decomposed
into the sum of the components $\eta_r$, $\eta_\varphi$, $\eta_z$, and $\eta_{\mathrm{mix}}$, which characterize the contributions of the field gradients in the radial, azimuthal, and axial directions as well as the interference between them.
We plot these components in Figs.~\ref{fig14} and \ref{fig15} as functions of the radial distance $r$ and the fiber radius $a$
for different guided modes. We observe from these figures that $\eta_z$ (dash-dotted blue curves), $\eta_r$ (dashed red curves), and $\eta_{\mathrm{mix}}$ (dash-dot-dotted magenta curves) are significant. Meanwhile, $\eta_\varphi$ (dotted green curves) is small except for the case of the HE$_{11}$ mode of a fiber with a small radius $a$. This feature is consistent with the fact that the expression for $\eta_\varphi$ in Eqs.~\eqref{a29} contains the factor $1/r^2$, which is small when $r$ and $a$ are large. With the help of an additional careful inspection of the dotted green curves in Figs.~\ref{fig14} and \ref{fig15}, we find that, among the contributions $\eta_\varphi^{\mathrm{HE11}}$, $\eta_\varphi^{\mathrm{TE01}}$, $\eta_\varphi^{\mathrm{TM01}}$, and $\eta_\varphi^{\mathrm{HE21}}$ of the gradient of the field in the azimuthal direction $\varphi$ to the  oscillator-strength enhancement factors for the HE$_{11}$, TE$_{01}$,  TM$_{01}$, and HE$_{21}$ modes, respectively, $\eta_\varphi^{\mathrm{HE21}}$ is the largest and $\eta_\varphi^{\mathrm{HE11}}$ is the smallest.
It is interesting to note that $\eta_\varphi^{\mathrm{HE11}}$, which corresponds to $l=1$, is smaller than $\eta_\varphi^{\mathrm{TE01}}$ and $\eta_\varphi^{\mathrm{TM01}}$, which correspond to $l=0$. The explanation is that the azimuthal gradient of the transverse component of the field in a quasicircularly polarized HE$_{11}$ mode is proportional to $ie_r-e_\varphi$ [see Eqs.~\eqref{a27}], which is small because $ie_r$ and $e_\varphi$ are real and have the same sign and comparable magnitudes  [see Eqs.~\eqref{g11}].

\begin{figure}[tbh]
\begin{center}
 \includegraphics{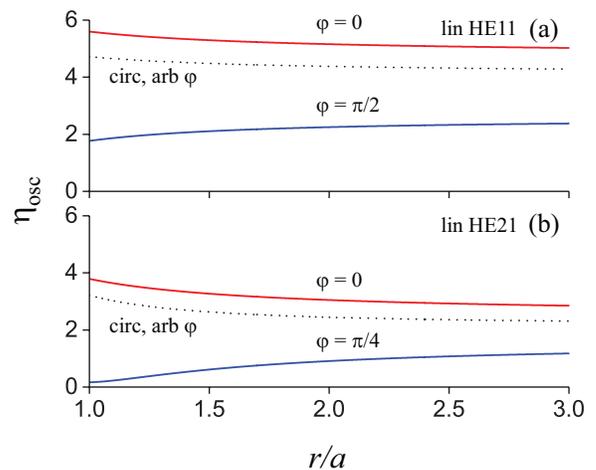}
 \end{center}
\caption{(Color online) Oscillator-strength enhancement factors $\eta_{\mathrm{osc}}$ for the quasilinearly polarized HE$_{11}$ and HE$_{21}$ modes
as functions of the radial distance $r$ at different azimuthal angles $\varphi$. The orientation angle of the quasilinear polarization axis is $\varphi_{\mathrm{pol}}=0$ and the quantization axis is arbitrary. Other parameters are as for Fig.~\ref{fig2}. 
For comparison, the results for the corresponding quasicircularly polarized hybrid modes are shown by the dotted black curves.}
\label{fig16}
\end{figure}

\begin{figure}[tbh]
\begin{center}
 \includegraphics{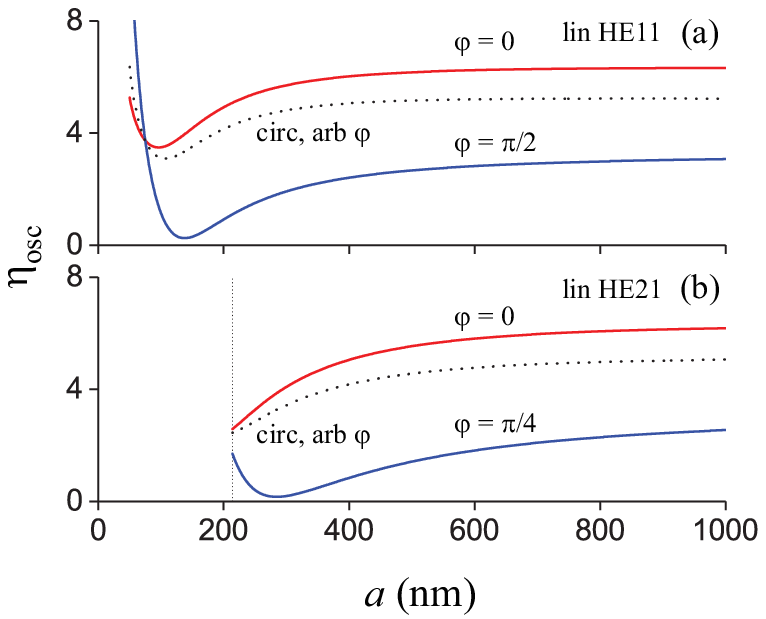}
 \end{center}
\caption{(Color online) 
Oscillator-strength enhancement factors $\eta_{\mathrm{osc}}$ for the quasilinearly polarized HE$_{11}$ and HE$_{21}$ modes as functions of the fiber radius $a$. The atom is positioned on the fiber surface at different azimuthal angles $\varphi$. 
The orientation angle of the quasilinear polarization axis is $\varphi_{\mathrm{pol}}=0$ and the quantization axis is arbitrary. Other parameters are as for Fig.~\ref{fig2}. The vertical dotted line indicates the position of the cutoff for the HE$_{21}$ mode. For comparison, the results for the corresponding quasicircularly polarized hybrid modes are shown by the dotted black curves.
}
\label{fig17}
\end{figure}

Due to the summation over transitions with different magnetic quantum numbers and the cylindrical symmetry of the field in a 
quasicircularly polarized hybrid mode, the oscillator strength $f_{FF'}$ and the enhancement factor $\eta_{\mathrm{osc}}$ for such a mode do not depend on the azimuthal position $\varphi$ of the atom in the fiber transverse plane. For the field in a quasilinearly polarized hybrid mode, since the cylindrical symmetry is broken, $f_{FF'}$ and $\eta_{\mathrm{osc}}$ vary with varying $\varphi$. We plot in Figs.~\ref{fig16} and \ref{fig17} the dependencies of $\eta_{\mathrm{osc}}$ for the quasilinearly polarized HE$_{11}$ and HE$_{21}$ modes on the radial distance $r$ and the fiber radius $a$ for different azimuthal angles $\varphi$. We observe from the figures that, depending on $\varphi$, the factor $\eta_{\mathrm{osc}}$ for a quasilinearly polarized hybrid mode may decrease or increase with increasing distance $r$, may be larger or smaller than that for the corresponding quasicircularly polarized hybrid mode, and may have a minimum in the dependence on the fiber radius $a$. Figure \ref{fig16} shows that $\eta_{\mathrm{osc}}$ varies slowly in the radial direction. Comparison between the curves for different
azimuthal angles in Figs.~\ref{fig16} and \ref{fig17} indicates that $\eta_{\mathrm{osc}}$ for quasilinearly polarized modes varies significantly in the azimuthal direction. 

\begin{figure}[tbh]
\begin{center}
 \includegraphics{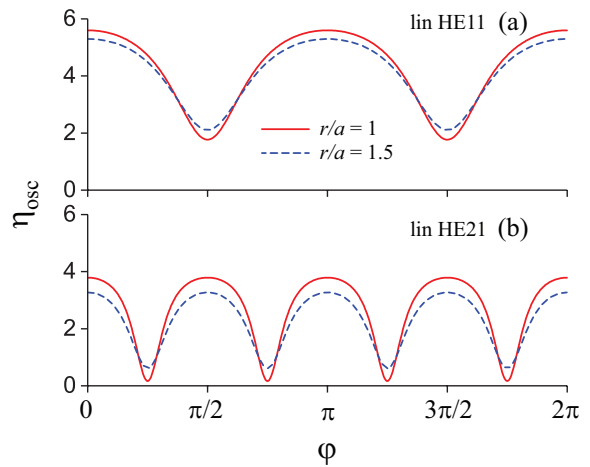}
 \end{center}
\caption{(Color online) Oscillator-strength enhancement factors $\eta_{\mathrm{osc}}$ for the quasilinearly polarized HE$_{11}$ and HE$_{21}$ modes as  functions of the azimuthal angle $\varphi$ for the position of the atom in the fiber cross-sectional plane. The orientation angle of the quasilinear polarization axis is $\varphi_{\mathrm{pol}}=0$ and the quantization axis is arbitrary. Other parameters are as for Fig.~\ref{fig2}. 
}
\label{fig18}
\end{figure}

\begin{figure}[tbh]
\begin{center}
 \includegraphics{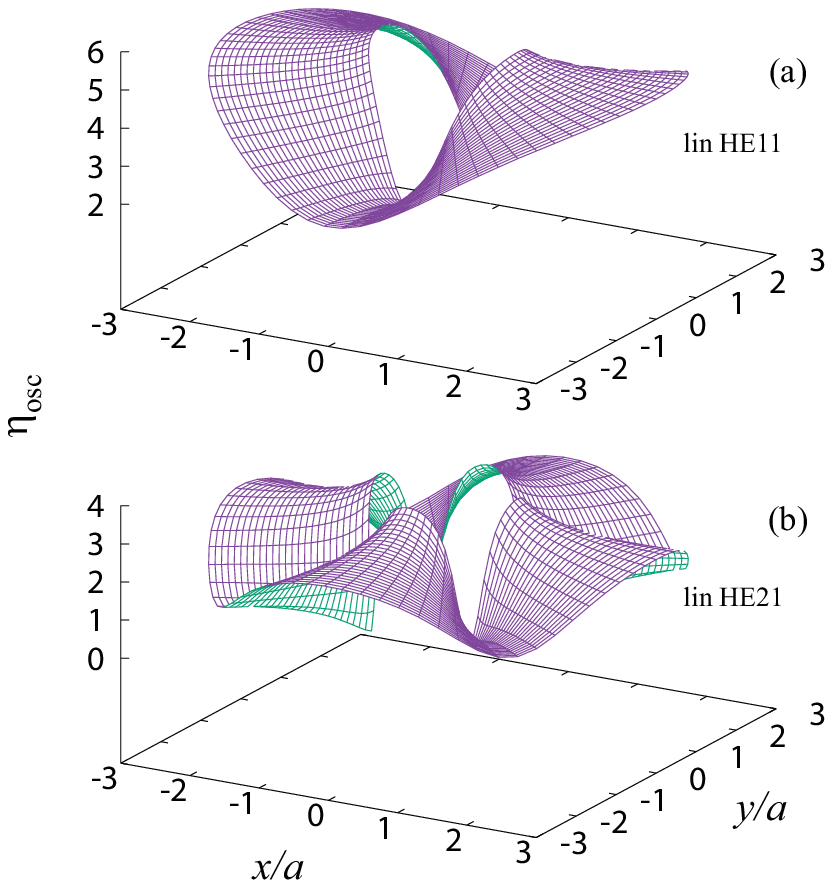}
 \end{center}
\caption{(Color online) Oscillator-strength enhancement factors $\eta_{\mathrm{osc}}$ for the quasilinearly polarized HE$_{11}$ and HE$_{21}$ modes as  functions of the position of the atom in the fiber cross-sectional plane. The orientation angle of the quasilinear polarization axis is $\varphi_{\mathrm{pol}}=0$ and the quantization axis is arbitrary. Other parameters are as for Fig.~\ref{fig2}. 
}
\label{fig19}
\end{figure}

In order to get a better view of the spatial profiles of the enhancement factor $\eta_{\mathrm{osc}}$ for quasilinearly polarized hybrid modes, we plot in Figs.~\ref{fig18} and \ref{fig19} this factor as a function of the azimuthal angle $\varphi$ and as a function of the Cartesian coordinates $x$ and $y$ of the position of the atom in the fiber cross-sectional plane. The figures show that $\eta_{\mathrm{osc}}$ for quasilinearly polarized hybrid modes varies significantly in the azimuthal direction but slightly in the radial direction, and is relatively large or small along the major or minor symmetry axes of the modes, respectively.

\section{Conclusion and discussion}
\label{sec:summary}

In this work, we have studied the electric quadrupole interaction of an alkali-metal atom with guided light in the fundamental and higher-order modes of a vacuum-clad ultrathin optical fiber. We have calculated the quadrupole Rabi frequency, the quadrupole oscillator strength, and their enhancement factors. In the example of a rubidium-87 atom, we have studied the dependencies of the Rabi frequency on the  quantum numbers of the transition, the mode type, the phase circulation direction, the propagation direction, the orientation of the quantization axis, the position of the atom, and the fiber radius. We have found that the rms quadrupole Rabi frequency and the quadrupole oscillator strength are enhanced by the effect of the fiber on the gradient of the field amplitude. With increasing radial distance, the rms Rabi frequency reduces quickly but the oscillator strength varies slowly. The enhancement factors of the rms Rabi frequency and the oscillator strength do not depend on any characteristics of the internal atomic states except for the atomic transition frequency. These factors are determined by the normalized spatial variations of the mode profile function at the atomic transition frequency. Like the oscillator strength, its enhancement factor $\eta_{\mathrm{osc}}$ varies slowly with increasing distance from the atom to the fiber surface. Due to this fact, the factor $\eta_{\mathrm{osc}}$ can be significant even when the atom is far away from the fiber. We have found that, in the case where the atom is positioned on the fiber surface, the oscillator strength for the quasicircularly polarized fundamental mode HE$_{11}$ has a local minimum at the fiber radius $a\simeq 107$ nm. Meanwhile, for quasicircularly polarized higher-order  hybrid modes, TE modes, and TM modes, the oscillator strength monotonically increases with increasing $a$. In the region $a<498.2$ nm, the oscillator strength for the quasicircularly polarized HE$_{11}$ mode is larger than that for quasicircularly polarized higher-order  hybrid modes, TE modes, and TM modes. We have shown that, depending on the azimuthal position of the atom, the enhancement factor $\eta_{\mathrm{osc}}$ for a quasilinearly polarized hybrid mode may decrease or increase with increasing distance, and may be larger or smaller than that for the corresponding quasicircularly polarized hybrid mode. We have found that the factor $\eta_{\mathrm{osc}}$ for quasilinearly polarized hybrid modes varies significantly in the azimuthal direction, and is relatively large or small along the major or minor symmetry axes of the modes, respectively. 

Our results may find application in future research on probing electric quadrupole transitions of atoms, molecules, and particles using the fundamental and higher-order modes of ultrathin optical fibers. Direct access to electric quadrupole transitions might be beneficial for fiber-based optical clocks \cite{Katori14}. A photon in a higher-order hybrid mode may have significant orbital angular momentum in addition to spin angular momentum. Therefore, our results on the enhanced electric quadrupole interaction of an atom with guided light might lead to an efficient way for transferring more than one quantum of angular momentum per photon to the internal degrees of freedom of the atom \cite{Kaler16,BDeb14}. Furthermore, the particular atomic transition addressed in this article allows one to prepare a rubidium atom in the excited state $4D_{5/2}$. The only dipole-allowed decay of this state to the ground state is via the intermediate level $5P_{3/2}$, by cascaded emission of two photons at 1530 nm and 780 nm. The emitted photons are correlated and can be entangled \cite{Kuzmich06,Roy17}. This opens up the possibility to develop a fiber-based source of entangled photon pairs at wavelengths relevant to telecom and atomic references.

\begin{acknowledgments}
We acknowledge support for this work from the Okinawa Institute of Science and Technology Graduate University.
\end{acknowledgments}


\appendix

\section{Matrix elements of the quadrupole tensor operators}
\label{sec:tensor}

We introduce the notations
\begin{equation}\label{t1}
\begin{split}
x_{-1}^{(1)}&=\frac{x_1-ix_2}{\sqrt2},\\
x_0^{(1)}&=x_3,\\
x_{1}^{(1)}&=-\frac{x_1+ix_2}{\sqrt2}
\end{split}
\end{equation}
for the spherical tensor components of the position vector $\mathbf{x}=(x_1,x_2,x_3)$. In terms of these components, we have
\begin{equation}\label{t2}
\begin{split}
x_{1}&=\frac{x_{-1}^{(1)}-x_1^{(1)}}{\sqrt2},\\
x_2&=i\frac{x_{-1}^{(1)}+x_1^{(1)}}{\sqrt2},\\
x_3&=x_0^{(1)}.
\end{split}
\end{equation}
We can write
\begin{equation}\label{t3}
x_i=\sum_q u_i^{(q)}x_q^{(1)},
\end{equation}
where $u_i^{(q)}$ with $i=1,2,3$ are the components of the spherical basis vectors $\mathbf{u}^{(q)}$ in the Cartesian coordinate system $\{x_1,x_2,x_3\}$.
The expressions for the vectors $\mathbf{u}^{(q)}$ are 
\begin{equation}\label{t4}
\begin{split}
\mathbf{u}^{(-1)}&=\frac{1}{\sqrt2}(1,i,0),\\
\mathbf{u}^{(0)}&=(0,0,1),\\
\mathbf{u}^{(1)}&=-\frac{1}{\sqrt2}(1,-i,0).
\end{split}
\end{equation}
We note that $\mathbf{u}^{(q)*}=(-1)^q \mathbf{u}^{(-q)}$, $\mathbf{u}^{(q)}\cdot\mathbf{u}^{(q')*}=\delta_{qq'}$,
and $\sum_q u_i^{(q)}u_j^{(q)*}=\delta_{ij}$.

It follows from Eq.~\eqref{t3} that
\begin{equation}\label{t5}
x_ix_j=\sum_{M_1M_2} u_i^{(M_1)}u_j^{(M_2)} x_{M_1}^{(1)}x_{M_2}^{(1)}.
\end{equation}
In order to calculate the direct product $x_{M_1}^{(1)}x_{M_2}^{(1)}$, we use the formula \cite{tensor}
\begin{equation}\label{t6}
x_{M_1}^{(1)}x_{M_2}^{(1)}=\sum_{Kq} T_{q}^{(K)}
(-1)^{q}\sqrt{2K+1}\begin{pmatrix}1&1&K\\M_1&M_2&-q\end{pmatrix},
\end{equation}
where $T_{q}^{(K)}$ with $q=-K,\dots,K$ are the tensor elements of the irreducible tensor products $T^{(K)}=[x^{(1)}\otimes x^{(1)}]^{(K)}$ of rank $K=0,1,2$. 
The expression for $T_{q}^{(K)}$ is
\begin{equation}\label{t7}
T_{q}^{(K)}=(-1)^{q}\sqrt{2K+1}\sum_{q_1q_2} 
\begin{pmatrix}1&1&K\\q_1&q_2&-q\end{pmatrix}
x^{(1)}_{q_1}x^{(1)}_{q_2}.
\end{equation} 
We can show that
\begin{equation}\label{t8}
\begin{split}
T_{0}^{(0)}&=-\frac{x_1^2+x_2^2+x_3^2}{\sqrt3},\\
T_{q}^{(1)}&=0,
\end{split}
\end{equation}
and
\begin{eqnarray}\label{t9}
T_{0}^{(2)}&=&\frac{2x_3^2-x_1^2-x_2^2}{\sqrt6},
\nonumber\\
T_{1}^{(2)}&=&-x_3(x_1+ix_2),
\nonumber\\
T_{-1}^{(2)}&=&x_3(x_1-ix_2),
\nonumber\\
T_{2}^{(2)}&=&\frac{1}{2}(x_1+ix_2)^2,
\nonumber\\
T_{-2}^{(2)}&=&\frac{1}{2}(x_1-ix_2)^2.
\end{eqnarray}
Note that $T_{0}^{(0)}=-R^2/\sqrt3$ and $T_{q}^{(2)}=2(2\pi/15)^{1/2} R^2Y_{2q}(\theta,\varphi)$, where $Y_{lq}(\theta,\varphi)$ are spherical harmonics
with $\theta$ and $\varphi$ being spherical angles.

We insert Eq.~\eqref{t6} into Eq.~\eqref{t5} and use Eq.~\eqref{a1}. Then, we obtain
\begin{equation}\label{t10}
Q_{ij}\equiv e(3x_ix_j-R^2\delta_{ij})=3e\sum_{q} u_{ij}^{(q)} T_{q}^{(2)},
\end{equation}
where
\begin{equation}\label{t11}
u_{ij}^{(q)}=(-1)^{q}\sqrt{5}\sum_{M_1M_2} u_i^{(M_1)}u_j^{(M_2)} 
\begin{pmatrix}1&1&2\\M_1&M_2&-q\end{pmatrix}.
\end{equation}
The explicit expressions for the tensors $u_{ij}^{(q)}$ are
\begin{equation}\label{t12}
\begin{split}
u_{ij}^{(2)}&=\frac{1}{2}\begin{pmatrix} 1&-i&0\\-i&-1&0\\0&0&0\end{pmatrix},\\
u_{ij}^{(1)}&=\frac{1}{2}\begin{pmatrix} 0&0&-1\\0&0&i\\-1&i&0\end{pmatrix},\\
u_{ij}^{(0)}&=\frac{1}{\sqrt6}\begin{pmatrix} -1&0&0\\0&-1&0\\0&0&2\end{pmatrix},\\
u_{ij}^{(-1)}&=\frac{1}{2}\begin{pmatrix} 0&0&1\\0&0&i\\1&i&0\end{pmatrix},\\
u_{ij}^{(-2)}&=\frac{1}{2}\begin{pmatrix} 1&i&0\\i&-1&0\\0&0&0\end{pmatrix}.
\end{split}
\end{equation}
Note that $u_{ij}^{(q)}=u_{ji}^{(q)}$, $u_{ij}^{(q)*}=(-1)^q u_{ij}^{(-q)}$, $\sum_{ij}u_{ij}^{(q)}u_{ij}^{(q')*}=\delta_{qq'}$, and $\sum_i u_{ii}^{(q)}=0$. 
 
The matrix elements of the tensor $T_{q}^{(2)}$ can be calculated using the Wigner-Eckart theorem \cite{tensor} 
\begin{multline}\label{t13}
\langle n'F'M'|T_{q}^{(2)}|nFM\rangle=\\ 
(-1)^{F'-M'}
\begin{pmatrix}F' &2 &F \\-M' & q& M\end{pmatrix}
\langle n'F'\|T^{(2)}\|nF\rangle.
\end{multline}
The invariant factor $\langle n'F' \| T^{(2)}\|nF \rangle$ is a reduced matrix element.
The selection rules for $F$ and $F'$ are $|F'-F|\le 2\le F'+F$.
The selection rules for $M$ and $M'$ are $|M'-M|\le 2$ and $M'-M=q$.
When we use Eqs.~\eqref{t10} and \eqref{t13}, we obtain \cite{James1998}
\begin{eqnarray}\label{t14}
\lefteqn{\langle n'F'M'|Q_{ij}|nFM\rangle=3e u_{ij}^{(M'-M)}(-1)^{F'-M'}}\nonumber\\
&&\mbox{}\times
\begin{pmatrix}F' &2 &F \\-M' & M'-M& M\end{pmatrix}
\langle n'F'\|T^{(2)}\|nF\rangle.\qquad
\end{eqnarray}

\section{Quadrupole interaction of an atom with a plane-wave light field in free space}
\label{sec:Rabi}

Assume that the field is a plane wave $\boldsymbol{\mathcal{E}}=\mathcal{E}\boldsymbol{\varepsilon}e^{i\mathbf{k}\cdot\mathbf{x}}$ in free space, where
$\mathcal{E}$ is the amplitude, $\mathbf{k}$ is the wave vector, and $\boldsymbol{\varepsilon}$ is the polarization vector.
In this case, the rms Rabi frequency $\bar{\Omega}_{FF'}^{(0)}$ is found from Eq.~\eqref{a10} to be
\begin{equation}\label{r1}
\bar{\Omega}_{FF'}^{(0)2}=\frac{e^2|\mathcal{E}|^2}{80\hbar^2}|\langle n'F'\|T^{(2)}\|nF\rangle|^2
\sum_q\Big|\sum_{ij}u_{ij}^{(q)}k_i\varepsilon_j\Big|^2.
\end{equation}

Without loss of generality, we assume that the field propagates along the $x_3$ direction and is linearly polarized along the $x_1$ direction.
Then, we have $\mathbf{k}=(0,0,k)$ and $\boldsymbol{\varepsilon}=(1,0,0)$ in the Cartesian coordinate system $\{x_1,x_2,x_3\}$. These expressions lead to
$k_i=k\delta_{i,3}$ and $\varepsilon_j=\delta_{j,1}$. Then, Eq.~\eqref{r1} gives
\begin{equation}\label{r2}
\bar{\Omega}_{FF'}^{(0)2}=\frac{e^2k^2|\mathcal{E}|^2}{80\hbar^2}|\langle n'F'\|T^{(2)}\|nF\rangle|^2\sum_{q}|u_{31}^{(q)}|^2.
\end{equation}
From Eqs.~\eqref{t12}, we find $\sum_{q}|u_{31}^{(q)}|^2=1/2$. Hence, we obtain 
\begin{equation}\label{r3}
\bar{\Omega}_{FF'}^{(0)2}=\frac{e^2k^2|\mathcal{E}|^2}{160\hbar^2}|\langle n'F'\|T^{(2)}\|nF\rangle|^2.
\end{equation}

The oscillator strength $f_{FF'}^{(0)}$ is related to the rms Rabi frequency $\bar{\Omega}_{FF'}^{(0)}$ via the formula \eqref{a11}.
With the help of this formula, we find
\begin{equation}\label{r4}
f_{FF'}^{(0)}=\frac{m_e\omega_0^3}{20\hbar c^2}\frac{|\langle n'F'\|T^{(2)}\|nF\rangle|^2}{2F+1}.
\end{equation}
The oscillator strength $f_{JJ'}^{(0)}$ of the transition from a lower fine-structure level $|nJ\rangle$ to an upper fine-structure level $|n'J'\rangle$
of the atom in free space may be obtained by summing up $f_{FF'}^{(0)}$ over all values of $F'$. The result is \cite{James1998,Freedhoff1989,Tojo2005b}  
\begin{equation}\label{r5}
f_{JJ'}^{(0)}=\frac{m_e\omega_0^3}{20\hbar c^2}\frac{|\langle n'J'\|T^{(2)}\|nJ\rangle|^2}{2J+1}.
\end{equation}

The rate $\gamma_{F'F}$ of quadrupole spontaneous emission from an upper hyperfine-structure  level $|n'F'\rangle$ to a lower hyperfine-structure level $|nF\rangle$ of the atom in free space is related to the oscillator strength $f_{FF'}^{(0)}$ as
\begin{equation}\label{r6}
\gamma_{F'F}^{(0)}=\frac{e^2\omega_0^2}{2\pi\epsilon_0 m_ec^3}\frac{2F+1}{2F'+1}f_{FF'}^{(0)}.
\end{equation}
Hence, we find
\begin{equation}\label{r7}
\gamma_{F'F}^{(0)}=\frac{e^2\omega_0^5}{40\pi\epsilon_0\hbar c^5}
\frac{|\langle n'F'\|T^{(2)}\|nF\rangle|^2}{2F'+1}.
\end{equation}
The rate $\gamma_{J'J}^{(0)}$ of quadrupole spontaneous emission from an upper fine-structure level $|n'J'\rangle$ to a lower fine-structure level $|nJ\rangle$ of the atom in free space may be obtained by summing up $\gamma_{F'F}^{(0)}$ over all values of $F$. The result is \cite{James1998,Freedhoff1989,Tojo2005b} 
\begin{equation}\label{r8}
\gamma_{J'J}^{(0)}=\frac{e^2\omega_0^5}{40\pi\epsilon_0\hbar c^5}
\frac{|\langle n'J'\|T^{(2)}\|nJ\rangle|^2}{2J'+1}.
\end{equation}
We have the relation
\begin{equation}\label{r9}
\gamma_{J'J}^{(0)}=\frac{e^2\omega_0^2}{2\pi\epsilon_0 m_ec^3}\frac{2J+1}{2J'+1}f_{JJ'}^{(0)}.
\end{equation}

It follows from Eq.~\eqref{a6} that the relations between $\gamma_{F'F}^{(0)}$ and $\gamma_{J'J}^{(0)}$ and between $f_{FF'}^{(0)}$ and $f_{JJ'}^{(0)}$ are
\cite{Tojo2004,Tojo2005a,error}
\begin{equation}\label{r10}
\begin{split}
\gamma_{F'F}^{(0)}&=(2F+1)(2J'+1)\begin{Bmatrix}F' &2 &F \\J & I& J'\end{Bmatrix}^2\gamma_{J'J}^{(0)},\\
f_{FF'}^{(0)}&=(2F'+1)(2J+1)\begin{Bmatrix}F' &2 &F \\J & I& J'\end{Bmatrix}^2 f_{JJ'}^{(0)}.
\end{split}
\end{equation}

\section{Guided modes of a step-index fiber}
\label{sec:guided}

Consider the model of a step-index fiber that is a dielectric cylinder of radius $a$ and refractive index $n_1$ and is surrounded by an infinite background medium of refractive index $n_2$, where $n_2<n_1$. 
For a guided light field of frequency $\omega$ (free-space wavelength $\lambda=2\pi c/\omega$ and free-space wave number $k=\omega/c$), the propagation constant $\beta$ is determined by the fiber eigenvalue equation \cite{fiber books}
\begin{eqnarray}\label{g1}
\lefteqn{\bigg[\frac{J_{l}'(ha)}{haJ_{l}(ha)}
+\frac{K_{l}'(qa)}{qaK_{l}(qa)}
\bigg]\bigg[\frac{n_{1}^2J_{l}'(ha)}{haJ_{l}(ha)}
+\frac{n_{2}^2K_{l}'(qa)}{qaK_{l}(qa)}
\bigg]}\nonumber\\&&\mbox{}\qquad\qquad\qquad\qquad
=l^2\left(\frac{1}{h^2a^2}+\frac{1}{q^2a^2}\right)^2\frac{\beta^2}{k^2}.
\end{eqnarray}
Here, we have introduced the parameters $h=(n_1^2k^2-\beta^2)^{1/2}$ and $q=(\beta^2-n_2^2k^2)^{1/2}$, which characterize the scales of the spatial variations of the field inside and outside the fiber, respectively. The integer index $l=0,1,2,\dots$ is the azimuthal mode order, which determines the helical phasefront and the associated phase gradient in the fiber transverse plane. 
The notations $J_l$ and $K_l$ stand for the Bessel functions of the first kind and the modified Bessel functions of the second kind, respectively. 
The notations $J'_l(x)$ and $K'_l(x)$ stand for the derivatives of $J_l(x)$ and $K_l(x)$ with respect to the argument $x$.
We note that the fiber eigenvalue equation \eqref{g1} remains the same when we replace $\beta$ by $-\beta$
or $l$ by $-l$. 

For $l\geq 1$, the eigenvalue equation \eqref{g1} leads to hybrid HE and EH modes \cite{fiber books}. The eigenvalue equation is given, 
for HE modes, as
\begin{equation}\label{g2}
\frac{J_{l-1}(ha)}{haJ_{l}(ha)}=-\frac{n_{1}^2+n_{2}^2}{2n_{1}^2}\frac{K'_{l}(qa)}{qaK_{l}(qa)}+
\frac{l}{h^2a^2}-\mathcal{R}
\end{equation}
and, for EH modes, as
\begin{equation}\label{g3}
\frac{J_{l-1}(ha)}{haJ_{l}(ha)}=-\frac{n_{1}^2+n_{2}^2}{2n_{1}^2}\frac{K'_{l}(qa)}{qaK_{l}(qa)}+
\frac{l}{h^2a^2}+\mathcal{R}.
\end{equation}
Here, we have introduced the notation
\begin{equation}\label{g4}
\begin{split}
\mathcal{R}&=\bigg[\bigg(\frac{n_{1}^2-n_{2}^2}{2n_{1}^2}\bigg)^2\bigg(\frac{K'_{l}(qa)}{qaK_{l}(qa)}\bigg)^2\\
&\quad +\bigg(\frac{l\beta}{n_{1}k}\bigg)^2\bigg(\frac{1}{q^2a^2}+\frac{1}{h^2a^2}\bigg)^2\bigg]^{1/2}.
\end{split}
\end{equation}
We label HE and EH modes as HE$_{lm}$ and EH$_{lm}$, respectively, where $l=1,2,\dots$ and $m=1,2,\dots$ are the azimuthal and radial mode orders, respectively. 
Here, the radial mode order $m$ implies that the HE$_{lm}$ or EH$_{lm}$ mode is the $m$th solution to the corresponding eigenvalue equation \eqref{g2} or \eqref{g3}, respectively.

For $l=0$, the eigenvalue equation \eqref{g1} leads to TE and TM modes \cite{fiber books}. The eigenvalue equation is given, for TE modes, as
\begin{eqnarray}\label{g5}
\frac{J_{1}(ha)}{haJ_{0}(ha)}=-\frac{K_{1}(qa)}{qaK_{0}(qa)}
\end{eqnarray}
and, for TM modes, as 
\begin{eqnarray}\label{g6}
\frac{J_{1}(ha)}{haJ_{0}(ha)}=-\frac{n_2^2}{n_1^2}\frac{K_{1}(qa)}{qaK_{0}(qa)}.
\end{eqnarray}
We label TE and TM modes as TE$_{0m}$ and TM$_{0m}$, respectively, where $m=1,2,\dots$ is the radial mode order. The subscript 0 implies that the azimuthal mode order of TE and TM modes is $l=0$.
The radial mode order $m$ implies that the TE$_{0m}$ or TM$_{0m}$ mode is the $m$th solution to the corresponding eigenvalue equation \eqref{g5} or \eqref{g6}, respectively.

According to \cite{fiber books}, the fiber size parameter $V$ is defined as $V=ka\sqrt{n_1^2-n_2^2}$.
The cutoff values $V_c$ for HE$_{1m}$ modes are determined as solutions to the equation $J_1(V_c)=0$. 
For HE$_{lm}$ modes with $l=2,3,\dots$, the cutoff values are obtained as nonzero solutions to the equation $(n_1^2/n_2^2+1)(l-1)J_{l-1}(V_c)=V_cJ_l(V_c)$. The cutoff values $V_c$ for EH$_{lm}$ modes, where $l=1,2,\dots$, are determined as nonzero solutions to the equation $J_l(V_c)=0$. 
For TE$_{0m}$ and TM$_{0m}$ modes, the cutoff values $V_c$ are obtained as solutions to the equation $J_0(V_c)=0$. 

The electric component of the field can be presented in the form
\begin{equation}\label{g7}
\mathbf{E}=\frac{1}{2}\boldsymbol{\mathcal{E}}e^{-i\omega t}+\mathrm{c.c.},
\end{equation}
where $\boldsymbol{\mathcal{E}}$ is the amplitude.
For a guided mode with a propagation constant $\beta$ and an azimuthal mode order $l$, we can write 
\begin{equation}\label{g8}
\boldsymbol{\mathcal{E}}=\mathbf{e}e^{i\beta z+il\varphi},
\end{equation}
where $\mathbf{e}$ is the mode profile function.
In order to construct the profile functions of a complete set of guided modes, 
we allow $\beta$ and $l$ in Eq.~\eqref{g8} to take not only positive but also negative values.  
We decompose the vectorial function $\mathbf{e}$ into the radial, azimuthal and axial components denoted by the subscripts $r$, $\varphi$ and $z$, respectively. We summarize the expressions for the mode functions of quasicircularly polarized hybrid modes, TE modes, and TM modes in the below \cite{fiber books}.

\subsection{Quasicircularly polarized hybrid modes}

We consider quasicircularly polarized hybrid modes $N=$ HE$_{lm}$ or EH$_{lm}$.
It is convenient to introduce the parameter
\begin{equation}\label{g9}
s=l\left(\frac{1}{h^2a^2}+\frac{1}{q^2a^2}\right)\left[\frac{J_{l}'(ha)}{haJ_{l}(ha)}
+\frac{K_{l}'(qa)}{qaK_{l}(qa)} \right]^{-1}.
\end{equation}
Then, we find, for $r<a$,
\begin{eqnarray}\label{g10}
e_{r}&=& iA\frac{\beta}{2h}[(1-s)J_{l-1}(hr)-(1+s)J_{l+1}(hr)],\nonumber\\
e_{\varphi}&=& -A\frac{\beta}{2h}[(1-s)J_{l-1}(hr)+(1+s)J_{l+1}(hr)],\nonumber\\
e_{z}&=& AJ_{l}(hr), 
\end{eqnarray}
and, for $r>a$,
\begin{eqnarray}\label{g11}
e_{r}&=& iA\frac{\beta}{2q}\frac{J_{l}(ha)}{K_{l}(qa)}[(1-s)K_{l-1}(qr)+(1+s)K_{l+1}(qr)],\nonumber\\
e_{\varphi}&=&-A\frac{\beta}{2q}\frac{J_{l}(ha)}{K_{l}(qa)}[(1-s)K_{l-1}(qr)-(1+s)K_{l+1}(qr)],\nonumber\\
e_{z}& = & A\frac{J_{l}(ha)}{K_{l}(qa)}K_{l}(qr).
\end{eqnarray}
Here, the parameter $A$ is a constant that can be determined from the propagating power of the field.

\subsection{TE modes}

We consider transverse electric modes $N=$ TE$_{0m}$.
For $r<a$, we have
\begin{eqnarray}\label{g12}
e_{r}&=&0,\nonumber\\
e_{\varphi}&=& i\frac{\omega\mu_{0}}{h}AJ_{1}(hr),\nonumber\\
e_{z}& = & 0. 
\end{eqnarray}
For $r>a$, we have
\begin{eqnarray}\label{g13}
e_{r}&=&0,\nonumber\\
e_{\varphi}&=&-i\frac{\omega\mu_{0}}{q}\frac{J_{0}(ha)}{K_{0}(qa)}AK_{1}(qr),\nonumber\\
e_{z}& = & 0.
\end{eqnarray}

\subsection{TM modes}

We consider transverse magnetic modes $N=$ TM$_{0m}$.
For $r<a$, we have
\begin{eqnarray}\label{g14}
e_{r}&=&-i\frac{\beta}{h}AJ_{1}(hr),\nonumber\\
e_{\varphi}&=& 0,\nonumber\\
e_{z}& = & AJ_{0}(hr). 
\end{eqnarray}
For $r>a$, we have
\begin{eqnarray}\label{g15}
e_{r}&=&i\frac{\beta}{q}\frac{J_{0}(ha)}{K_{0}(qa)}AK_{1}(qr),\nonumber\\
e_{\varphi}&=&0,\nonumber\\
e_{z}& = & \frac{J_{0}(ha)}{K_{0}(qa)}AK_{0}(qr).
\end{eqnarray}


\end{document}